\pdfoutput=1
\documentclass[twocolumn,
               showpacs,
               preprintnumbers,
               nofootinbib,
               prd,
               superscriptaddress,
               10pt,
               notitlepage,
               aps]{revtex4-2}
               
\usepackage{graphicx,amssymb,amsmath,amsthm,amsfonts,epsfig,mathtools}

\usepackage[utf8]{inputenc}
\usepackage{graphicx}
\usepackage{dcolumn}
\usepackage{bm}
\usepackage{amsmath}
\usepackage{color}
\usepackage[dvipsnames]{xcolor}
\usepackage{hyperref}
\hypersetup{colorlinks=true, citecolor=MidnightBlue,linkcolor=CornflowerBlue, urlcolor=CornflowerBlue, linktocpage=true}
\usepackage{xfrac}
\usepackage{siunitx}

\newcommand{\beqn}{\begin{eqnarray}}
\newcommand{\eeqn}{\end{eqnarray}}
\newcommand{\beq}{\begin{equation}}
\newcommand{\eeq}{\end{equation}}

\def\mphi{m_{\phi}}

\def\pt{\tilde{p}}
\def\rt{\tilde{\rho}}

\def\tg{\tilde{g}}

\def\qt{\tilde{q}}

\def\meff{m_{\textrm{eff}}}

\begin{document}

\title{Constraining scalar-tensor theories using neutron star mass and radius measurements}

\begin{abstract}
We use neutron star mass and radius measurements to constrain the spontaneous scalarization phenomenon in scalar-tensor theories using Bayesian analysis. Neutron star structures in this scenario can be significantly different from the case of general relativity, which can be used to constrain the theory parameters. We utilize this idea to obtain lower bounds on the coupling parameter $\beta$ for the case of massless scalars. These constraints are currently weaker than the ones coming from binary observations, and they have relatively low precision due to the approximations in our method. Nevertheless, our results clearly demonstrate the power of the mass-radius data in testing gravity, and can be further improved with future observations. The picture is different for massive scalars, for which the same data is considerably less effective in constraining the theory parameters in an unexpected manner. We identify the main reason for this to be a large high-likelihood region in the parameter space where deviations from general relativity are relatively small. We hope this initial study to be an invitation to use neutron star structure measurements more commonly to test alternative theories in general.
\end{abstract}

\author{Semih Tuna}
\email{semih.tuna@columbia.edu}
\affiliation{Department of Physics, Columbia University, New York, New York 10027, USA}

\author{K{\i}van\c{c} \.I. \"Unl\"ut\"urk}
\email{kunluturk17@ku.edu.tr}
\affiliation{Department of Physics, Ko\c{c} University, \\
Rumelifeneri Yolu, 34450 Sariyer, Istanbul, Turkey}

\author{Fethi M. Ramazano\u{g}lu}
\email{framazanoglu@ku.edu.tr}
\affiliation{Department of Physics, Ko\c{c} University, \\
Rumelifeneri Yolu, 34450 Sariyer, Istanbul, Turkey}

\date{\today}
\maketitle

\section{Introduction}\label{intro}
Understanding the structure of neutron stars is a primary goal of both astrophysics and nuclear physics. On one hand, these objects are perfect tools to investigate the effects of extreme gravity~\cite{LIGOScientific:2018dkp,Barack:2018yly}, on the other, one can use the observational data for a better understanding of the behavior of nuclear matter~\cite{Lattimer:2012nd,Ozel:2016oaf,LIGOScientific:2018cki}.

Many aspects of nuclear matter are reflected in the relationship between the mass and radius of neutron stars. The radius of a neutron star is mainly determined by its mass and the nuclear matter equation of state
(EOS), in addition to some other factors such as spin~\cite{Shapiro:1983du}. This means simultaneous observations of neutron star masses and radii, i.e., mass-radius measurements, can be inverted to deduce the nuclear EOS. The fact that the EOS has not been computed from first principles with high accuracy due to the nonperturbative nature of quantum chromodynamics, especially for higher nucleon densities, makes this astrophysical strategy all the more valuable~\cite{Lattimer:2012nd, Ozel:2016oaf}.

Most of the work on the relation between neutron star mass-radius relationship and the nuclear EOS assumes general relativity (GR) as the governing theory of gravitation. However, deviations from GR could affect the star structure as well, hence, simultaneous mass-radius measurements can also be used to test alternative theories of gravitation. Distinguishing small deviations from GR is not usually possible due to the large uncertainties in the data, but theories that predict large deviations, in principle, can be constrained.

In this study, we will assess the ability of neutron star mass-radius data to constrain the parameter space of the so-called spontaneous scalarization phenomenon. Spontaneous scalarization is a scenario in scalar-tensor theories where neutron stars are naturally surrounded by scalar field clouds for certain values of compactness~\cite{PhysRevLett.70.2220}. These clouds grow due to a tachyonic instability in regions of spacetime with high energy densities, and lead to large deviations from GR for neutron stars that go through this process. The scalar field dies off away from the star, hence weak field tests of gravity are also satisfied~\cite{Will:2001LR}. This latter feature means that constraining spontaneous scalarization using weak-field tests is hard, which makes neutron star observations particularly valuable.

Mass-radius measurements obtained through X-ray data have been successfully used to constrain the EOS under the assumption of GR using Bayesian analysis~\cite{Ozel:2015fia}. Our approach in this paper will be similar, where we constrain the parameters of spontaneous scalarization theories rather than the EOS. Even though there are various different models of spontaneous scalarization, our emphasis will be on the original Damour-Esposito-Farèse (DEF) model~\cite{PhysRevLett.70.2220} and its extension to massive scalars~\cite{Ramazanoglu:2016kul}. We parametrize these theories with two parameters: the quadratic scalar coupling coefficient $\beta$ and the mass of the scalar $\mphi$.

We will first demonstrate that for the original DEF model ($\mphi=0$), neutron star mass-radius data can provide a lower bound for $\beta$: $\beta \gtrsim -15$. Our bounds are less stringent than the already existing ones obtained from binary star systems~\cite{2013Sci...340..448A,Freire:2012mg,Zhao:2022vig}, but this demonstrates the effectiveness of mass-radius data in constraining deviations from GR as a proof of principle. Furthermore, our bounds are likely to improve as more data is obtained using electromagnetic \cite{Bogdanov:2019ixe,Bogdanov:2019qjb,Bogdanov:2021yip} and gravitational wave observations~\cite{LIGOScientific:2018cki,Niu:2021nic}.

The effect of the scalar mass $\mphi$ is drastic on scalarization, and it renders the known bounds from binary observations ineffective due to the radical change in the far-field behavior~\cite{Ramazanoglu:2016kul}. This means that the $\beta$ parameter cannot be constrained for generic values of $\mphi \neq 0$. On the other hand, very large negative values of $\beta$ are known to significantly change the neutron star structure compared to the case of GR, and mass-radius observations can be naively expected to rule out at least this part of the parameter space. However, our results show that this is surprisingly not the case, and the current mass-radius data alone cannot put even loose lower bounds on $\beta$ independently of $\mphi$.

We identify the main reason for the difference in the effectiveness of Bayesian analysis between the massless and massive scalars to be the fact that $\mphi$ suppresses scalarization, hence deviations from GR, while increasingly negative values of $\beta$ facilitates it. This fact was known from previous studies~\cite{Ramazanoglu:2016kul}, but we discovered that the region where the star structure stays relatively close to that of GR on the $(\beta,\mphi)$ parameter space is quite large. The mass-radius curve can change continuously with the parameters to provide a better fit to the observations in this region, which extends to $\beta \ll -1$. Therefore, the marginal posterior likelihood is relatively independent of $\beta$ for many choices of the prior distribution, and we cannot provide clear bounds on this parameter. Constraining $\mphi$ is also not possible due to the difficulty in choosing a physically-motivated prior distribution as we shall discuss. This general picture is partially due to our current lack of strong bounds on the scalar field mass $\mphi$, hence it might change if constraints that are obtained from other measurement channels improve in the future, one possibility being novel observations of highly spinning black holes~\cite{Stott:2018opm,Stott:2020gjj}.

One ideally needs to reanalyze the X-ray optical data for spontaneously scalarized neutron stars (instead of stars in GR) to relate the observations to the star structure. The basic framework for this has been worked out in some cases~\cite{Silva:2018yxz,Hu:2021tyw}, but we do not yet have a complete dataset of neutron star masses and radii computed assuming this alternative theory as was done in GR~\cite{Ozel:2015fia}, and such a calculation is an involved project on its own. Instead, we use the mass-radius data obtained under the assumption of GR, together with some simplifying approximations about the spacetime of a scalarized neutron star, which we present in detail in Sec.~\ref{sec:mass_radius_curves}. We employ different approximations to assess the error introduced by them, and see that our main results stay qualitatively same in all cases. Overall, this means that obtaining precise bounds on the theory parameters of spontaneous scalarization models requires further work. However, the main conclusions of this initial study seem robust: mass-radius data can be used for constraining spontaneous scalarization for massless scalars, but it is not effective for massive scalars unless some independent constraint on the scalar mass is obtained.

We consider a specific spontaneous scalarization theory, however, our methods can be readily applied to any alternative theory of gravitation. Neutron star mass-radius data is especially effective at testing theories that predict relatively large changes in the star structure compared to GR, chief examples being more recent models of spontaneous scalarization that couple the scalars to curvature rather than matter~\cite{Doneva:2017bvd,Silva:2017uqg,Herdeiro:2018wub,Andreou:2019ikc}. We hope our work encourages the scientific community to utilize the neutron star structure measurements more effectively for studying deviations from GR.

In Sec.~\ref{spontaneous_scalarization}, we introduce spontaneous scalarization theories with massive fields, and discuss the qualitative dependence of neutron star structure on the parameters of the theory. In Sec.~\ref{methods}, we discuss the methods to compute the neutron star structures and the details of the Bayesian analysis. In Sec.~\ref{results}, we present the constraints we obtained on the parameter space of spontaneous scalarization. We finally discuss our results and possible future directions in Sec.~\ref{sec:conclusions}. Further details of the numerical computations are provided in the appendices.

\section{Scalarization trends in the $(\beta, \mphi)$ parameter space}\label{spontaneous_scalarization}
Spontaneous scalarization is a specific scenario in scalar-tensor theories which can be given by the action~\cite{PhysRevLett.70.2220, Ramazanoglu:2016kul}
\begin{align}\label{st_action}
  \frac{1}{16\pi} &\int d^4x \sqrt{-g}\ \bigg[R
 -2g^{\mu\nu}  \nabla_{\mu} \phi  \nabla_{\nu} \phi\
 -2 m_{\phi}^2 \phi^2 \bigg] \nonumber \\
 &+ S_\text{m} \left[f_\text{m}, \tg_{\mu \nu} \right] \ ,
\end{align}
where $\tg_{\mu\nu} = A^2(\phi) g_{\mu \nu}$ is the so-called Jordan frame metric, and $f_\text{m}$ represents any degrees of freedom other than $g_{\mu\nu}$ or $\phi$, e.g. matter in a neutron star. Even though many other models of scalarization have been introduced, we will be exclusively studying this theory. For a specific class of conformal scaling functions $A(\phi)$ specified below, this is the original Damour-Esposito-Farèse (DEF) model with the addition of the intrinsic mass of the scalar field, $\mphi$. 

Action~\eqref{st_action} can represent a general scalar-tensor theory; spontaneous scalarization occurs when the conformal scaling function $A(\phi)$ has a specific form such that the uniform $\phi=0$ configuration can be a solution of the theory, but not always a stable one. Arbitrarily small perturbations in $\phi$ start growing exponentially around some neutron stars and eventually settle to configurations where stars are surrounded by stable scalar clouds. The final ``scalarized'' solution typically features large deviations from GR due to the large amplitudes of the scalar field.

The underlying reason for the initial growth and the final stable configuration of the scalar field can be understood by studying its field equation 
\begin{align} \label{scalar_eom}
  \Box_g \phi &= \left( - 8 \pi A^4 \frac{d\left( \ln A \right)}{d(\phi^2)} \tilde{T} + m^2_\phi \right)\phi 
  = \meff^2\ \phi \ ,
\end{align}
where $\tilde{T}=\tilde{g}_{\mu\nu}\tilde{T}^{\mu\nu}$ denotes the trace of the stress-energy tensor in the Jordan frame, and we defined the effective mass
\begin{align} \label{meff}
 \meff^2 \equiv  -8 \pi A^4 \frac{d\left( \ln A \right)}{d(\phi^2)} \tilde{T} + m^2_\phi \ .
\end{align}
In short, the scalar behaves as if its mass is $ \meff$. Then, consider
\begin{align}
A=e^{\beta\phi^2/2}\ ,
\label{eq:conformal_function}
\end{align}
where $\beta$ is a constant, which is the primary case originally considered by Damour and Esposito-Farèse~\cite{PhysRevLett.70.2220}. For small values of $\phi$ we can linearize the equation of motion around $\phi=0$:
\begin{align} \label{meff_linear}
  \Box_g \phi = \meff^2\ \phi \ , \ 
 \meff^2 \approx  - 4 \pi \beta \tilde{T} + m^2_\phi \ .
\end{align}
For typical matter, even for neutron stars in many cases, $\tilde{T} = -\tilde{\rho}+3\tilde{p} \approx -\tilde{\rho} < 0$. This means for appropriate values of $\beta<0$, $\meff$ becomes imaginary, which indicates a tachyonic instability. In rough terms, a Fourier mode $e^{-i \omega t} e^{k_i x^i}$ has the dispersion relation $\omega \sim \sqrt{k_ik^i+ \meff^2}$. Hence, the frequencies for low enough wave numbers $k$ are imaginary, which leads to exponential growth in time rather than oscillation. This is the origin of the instability around $\phi=0$~\cite{Ramazanoglu:2016kul}.

The second main aspect of spontaneous scalarization is the fact that the instability is suppressed as the field grows, leading to a stable scalar cloud. Beyond the linearized approximation, the tachyonic contribution to the effective mass, $-4\pi A^4  \beta \tilde{T}$, diminishes as $\phi$ grows thanks to the $A(\phi \to \infty)$ behavior in Eq.~\eqref{eq:conformal_function} when $\beta<0$. The scalar is unstable around $\phi=0$, but it is also self-regulating for large scalar values.

The mechanism for the instability and its suppression do not have any direct references to neutron stars so far; they can occur for any form of matter. However, note that the effective mass can only be tachyonic  inside matter, and the highest wavelength (lowest $k$) Fourier mode we can fit inside a given matter distribution is restricted by its size. Hence, for a given $\beta$, there is a lower bound to the compactness of a star below which there is no scalarization~\cite{Ramazanoglu:2016kul}. For $A=e^{\beta \phi^2/2}$, order-of-unity values of $\beta$ can only lead to spontaneous scalarization for densities found inside neutron stars. The next most compact class of objects are white dwarfs that require $\beta \sim -10^3$ to scalarize, and we will not consider such values in this work.

The intrinsic mass of the scalar, $\mphi$, did not play any direct role in our explanation of spontaneous scalarization. This is because scalar mass actually makes scalarization harder and even impossible if it is above a certain value as can be seen in Eq.~\eqref{meff_linear}. For any given $\beta$, there is a high enough $\mphi$ which makes the effective squared mass positive for any conceivable density inside a neutron star. Scalar mass was indeed not considered in the original description of the DEF model~\cite{PhysRevLett.70.2220}, and was introduced later when massless theories were largely ruled out by more recent observations~\cite{Ramazanoglu:2016kul}. Scalar fields of static scalarized stars die off with a power of the distance from the star if the field is massless, as opposed to the much faster exponential decay for massive fields. As a result, massive scalars are not constrained by binary star observations when the Compton wavelength of the scalar is much smaller than the binary separation~\cite{Ramazanoglu:2016kul}, whereas most of the parameter space of the $\mphi=0$ case has been ruled out~\cite{2013Sci...340..448A,Freire:2012mg,Zhao:2022vig}. Since the binary observations are the main source of the bounds on scalarization parameters, the parameter space for $\mphi \neq 0$ is still largely viable.

The form of $A(\phi)$ is not crucial for the onset of instability, and any function with the behavior $A(\phi) = 1 +\beta \phi^2/2 + \ldots$ in its Taylor expansion leads to similar results. Because of this, we will only consider $A=e^{\beta \phi^2/2}$ of Eq.~\eqref{eq:conformal_function} in this work, which is also the original choice of DEF~\cite{PhysRevLett.70.2220}. However, the exact form of $A$ can lead to differences in the final stable scalarized configurations, an issue we will discuss in the conclusions.
{
\begin{figure}
\begin{center}
\includegraphics[width=.49\textwidth]{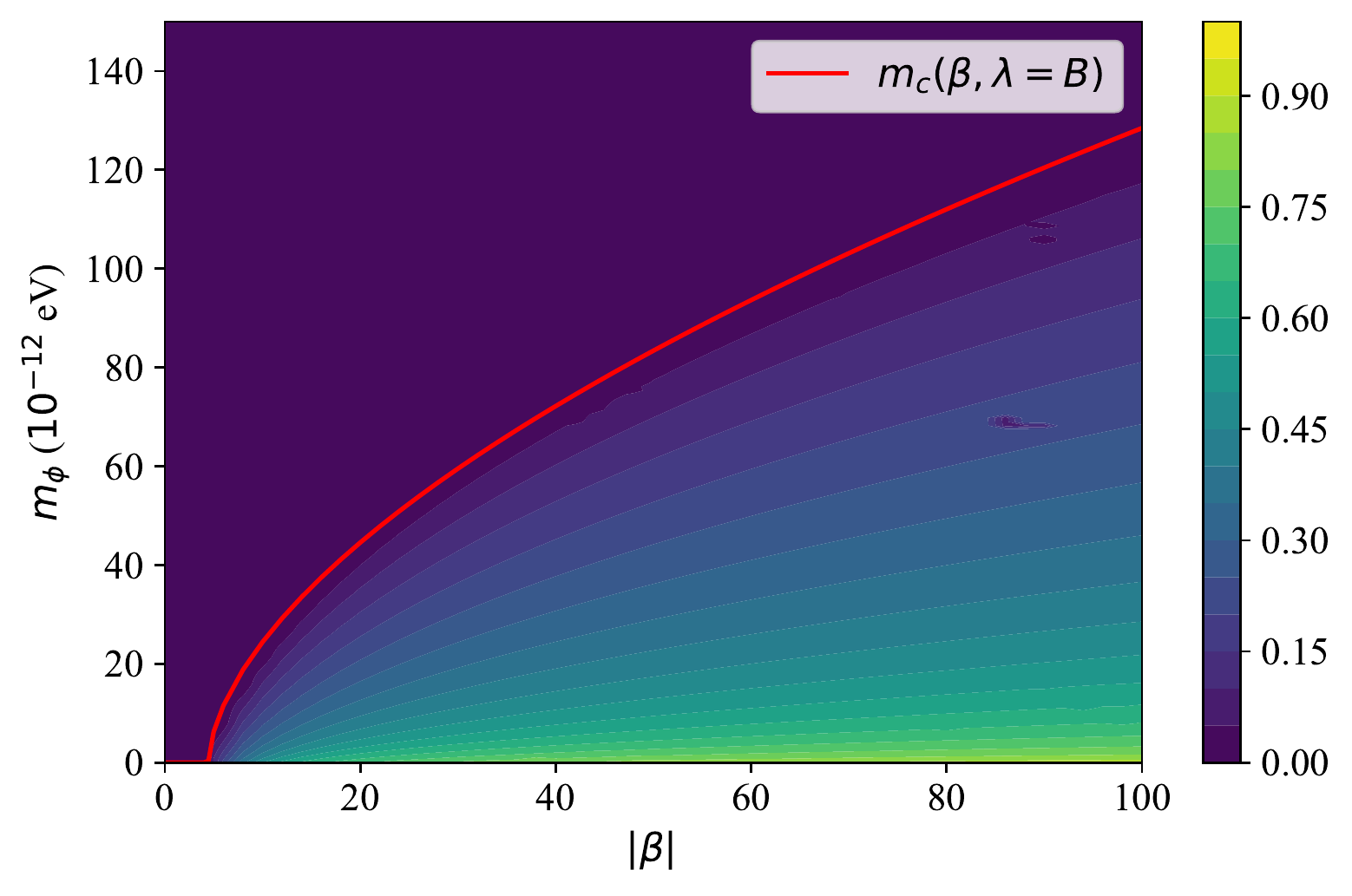}
\end{center}
\caption{
Strength of scalarization roughly measured by the maximum computed value of $1-A_c$ for given $(\beta,\mphi)$ and B EOS, $A_c$ being the value of the conformal scaling function at the center of the star. Stronger scalarization is achieved by higher $|\beta|$ and lower $\mphi$ for $\beta<0$. There is no scalarization above and to the left of the critical mass curve $m_c(\beta)$. We will refer to this part of the parameter space as the GR-equivalent region. There is also no scalarization for $\beta>\beta_c$ which satisfies $m_c(\beta_c)=0$. Scalarization is monotonically stronger with growing $\beta$ when $\mphi=0$, but the two parameters have opposite effects. The few spotty regions near the $|\beta|=90$ line are due to the failure of our numerical solver, which we explain further in Appendix~\ref{sec:interpolation}.}
\label{m_c_plot}
\end{figure}

In light of the above explanations, we will parametrize spontaneous scalarization with the two parameters $(\beta,\mphi)$ and the condition $\beta<0$ (see Sec.~\ref{sec:prior} for more on this condition). The strength of scalarization in a neutron star, hence deviations from GR, depends on these parameters as follows (also see Fig.~\ref{m_c_plot}).
\begin{enumerate}
\item Scalarization becomes more prominent with increasing $-\beta$, i.e. increasing $|\beta|$, since the tachyonic terms become predominant. Specifically, static neutron stars for $\beta=0$ are exactly those of GR, irrespective of $\mphi$.\footnote{Whether $\beta=0$ should be considered as GR is a matter of definition. There is no nonminimal coupling and no scalarization in this case, but the scalar field still exists in the theory and might have other observable signatures as we will see in Sec.~\ref{sec:mphi_constraint}. This case can be viewed as GR with a hitherto unknown scalar field.} Moreover, there is an upper limit to $\beta$ such that scalarization does not occur at all for $\beta>\beta_c$. The critical value $\beta_c$ slightly changes with EOS, but is known to be roughly around $-4.5$ for commonly studied cases~\cite{Ramazanoglu:2016kul,2013Sci...340..448A}.

\item Scalarization becomes less prominent with increasing $\mphi$. For any given $\beta$, there is a critical mass $m_c(\beta)$ such that neutron stars do not scalarize at all if $\mphi > m_c(\beta)$, where $m_c$ has a slight dependence on EOS, like $\beta_c$. This means our extended DEF theory is equivalent to GR in this part of the parameter space as far as mass-radius data is concerned. Since the square of the effective mass grows linearly with $\beta$, asymptotically
\begin{align}
m_c(\beta) \sim \sqrt{-\beta} \ .
\end{align}

\item The structure of scalarized neutron stars for small scalar field masses $\mphi \lesssim 10^{-14}$~eV is practically identical to those with $\mphi=0$, aside from a very small region around $\beta_c$ in the parameter space. This is because such a small value of $\mphi$ does not contribute significantly to the effective mass unless $\beta \approx \beta_c$.

\end{enumerate}
We will soon see that the effects of $|\beta|$ and $\mphi$ we explained mean there is a region of the parameter space where scalarization, and hence deviations from GR, are relatively low. This will make a drastic difference between massless and massive scalars for our results.

The amount of scalarization in general depends on the EOS. Even though the qualitative behavior of scalarization for different EOS is similar, the differences are large enough to consider a variety of them as an additional fitting parameter in our analysis. However, we should emphasize that our main aim is constraining the parameter space for spontaneous scalarization, and our consideration of EOS is mainly to analyze the basic aspects of the dependence of our results on this unknown factor.

\section{Computational Methods}\label{methods}
\subsection{Obtaining the mass-radius curves}
\label{sec:mass_radius_curves}
The essence of our work is utilizing the dependence of the mass and radius of a neutron star on $(\beta,\mphi)$ and the EOS. Therefore, we need to compute the scalarized neutron star structures to obtain the mass-radius curves, which is only possible with numerical methods.

For a static spherically symmetric star of perfect fluid matter, we use the metric ansatz
\begin{equation}\label{metric_ansatz}
g_{\mu\nu} dx^{\mu} dx^{\nu} = -e^{\nu(r)} dt^2 + \frac{dr^2}{1-2\mu(r)/r} + r^2 d\Omega^2\ 
\end{equation}
and the perfect fluid stress-energy tensor with respect to the metric $\tilde{g}_{\mu\nu}$
\begin{equation}
\tilde{T}^{\mu\nu}=(\rt+\pt)\tilde{u}^{\mu}\tilde{u}^{\nu}+\pt \tilde{g}^{\mu\nu} \ .
\end{equation}
The resulting modified Tolman-Oppenheimer-Volkov (TOV) equations are~\cite{Ramazanoglu:2016kul}
\begin{align} \label{tov}
 \mu' &= 4\pi r^2 A^4 \rt + \frac{1}{2}r(r-2\mu) \psi^2 + \frac{1}{2} r^2 \mphi^2 \phi^2 \nonumber \\
 \nu' &= r\psi^2 + \frac{r^2}{(r-2\mu)}\left[(8\pi A^4\pt-\mphi^2\phi^2)+2\frac{\mu}{r^3} \right] \nonumber \\
 \phi'&= \psi\\
 \psi' &(r-2\mu) = 4\pi r A^4 \left[\alpha(\rt-3\pt)+r\psi(\rt-\pt)\right] \nonumber \\
 &\ \ \ \ \ \ \ \ \ \ \ \ \  +\mphi^2 (r^2\phi^2\psi +r\phi)-2\psi(1-\mu/r) \nonumber \\
 \pt' &= -(\rt+\pt)\left( \nu'/2+\alpha \psi \right) \ .  \nonumber 
\end{align}
where $\alpha = d\left(\ln A\right)/d\phi=\beta \phi$, all variables are only functions of $r$, and $'$ denotes a derivative with respect to this coordinate. The system of equations closes given an EOS $\rt=\rt(\pt)$.

We know that $\mu(0)=\psi(0)=0$ due to regularity. Physical solutions also have to satisfy $\phi(\infty)=0$. Then, the main numerical problem is finding the value of $\phi_c \equiv \phi(0)$ which satisfies the physicality condition $\phi(\infty)=0$ for a given value of $\pt_c \equiv \pt(0)$, or a given star radius. We solve this boundary value problem using a relaxation method inspired by the recent work of Ref.~\cite{Rosca-Mead:2020bzt}, whose details are given in Appendix~\ref{app:mass_radius}. 

Mass-radius curves are obtained by computing many stars for each point on the parameter space, sample cases can be seen in Fig.~\ref{mr_curve_beta_low}. Details of obtaining these curves from individual star solutions, and repeating the process for all points on the parameter space requires care since the total computational cost can be prohibitive otherwise. Details of this procedure can be found in Appendix~\ref{sec:app_mr_curve}.
\begin{figure}
\includegraphics[width=.49\textwidth]{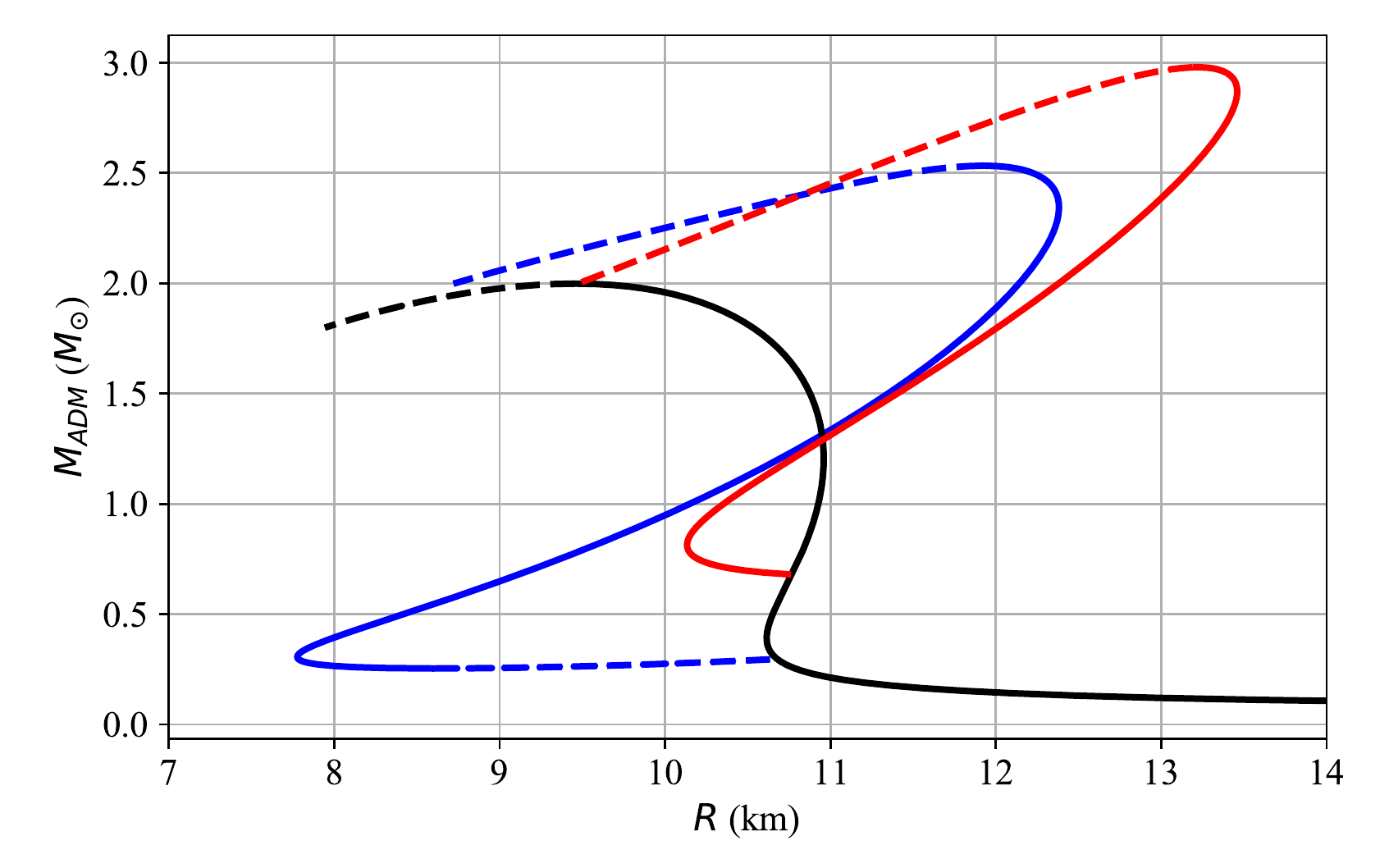}
\caption{
Mass-radius curves for GR (black), and two different cases of spontaneous scalarization with $\beta = -8$, $\mphi= 1.5 \times 10^{-12}$~eV (red) and $\beta=-34$, $\mphi= 3 \times 10^{-11}$~eV (blue). B EOS is used in all cases. Stable solutions are shown with solid lines, and unstable ones with dashed ones (see Appendix~\ref{sec:app_mr_curve}). Note that a large part of the stable portion of the GR curve is unstable to scalarization in the scalar-tensor theory, but this curve determines the mass-radius relationship when there are no stable scalarized neutron stars with a given mass, such as the low mass region on the lower left part of the figure. Once the maximal neutron star mass is reached, the solutions continually connected to this one to the left of the plot are always unstable (upper dashed parts). In addition, some $(\beta,\mphi)$ values present another unstable region at low star masses, where the scalarized branch deviates from the GR one for the first time (blue curve, lower dashed part).}
\label{mr_curve_beta_low}
\end{figure}

All the non-metric degrees of freedom, including the matter and electromagnetic waves relevant for the observations, minimally couple to the metric $\tg_{\mu\nu} = A^2(\phi) g_{\mu \nu}$, which can be put into the same form as $g_{\mu\nu}$ as
\begin{equation}
\tg_{\mu\nu} dx^{\mu} dx^{\nu} = -e^{\tilde{\nu}} dt^2 + \frac{d\tilde{r}^2}{1-2\tilde{\mu}/\tilde{r}} + \tilde{r}^2 d\Omega^2\ 
\end{equation}
where
\begin{align}
\tilde{r} &= A r\ \  \Rightarrow\ \
d\tilde{r} = A(1+\alpha \psi) dr \nonumber\\
\tilde{\nu} &= \nu + 2 \ln A \nonumber\\
\frac{d\tilde{r}^2}{1-2\tilde{\mu}/\tilde{r}} &= 
A^2\frac{dr^2}{1-2\mu/r}\\ \Rightarrow\ \ 
\tilde{\mu} &= (1+\alpha \psi)^2 A \mu - \frac{1}{2}A \alpha \psi (2+\alpha\psi) r
\end{align}
We will use the values associated with this metric in our Bayesian analysis.

Associating the neutron star structures computed as above to the data has some significant approximations built into it. Firstly, Ref.~\cite{Ozel:2015fia}, whose observational data we use, analyzes spinning neutron stars, which are not spherically symmetric. Spin modifies the structure as well as the amount of scalarization of a neutron star, specifically, spin tends to strengthen scalarization, in general leading to further deviations from GR~\cite{Yazadjiev:2016pcb}. Computation of the structure of spinning stars are numerically much costlier than the spherically symmetric ones, hence we will only consider spherical symmetry in this initial work at the cost of losing precision in our results. Incorporating rotation effects will be one of our goals in the future.

The second approximation we will use is due to the mass-radius data, that of Ref.~\cite{Ozel:2015fia}, which was obtained under the assumption of GR. $\tilde{\mu}(r)=\mu(r)$ is a constant outside the star in this case, which is also the stellar mass. This is not valid for spontaneous scalarization, where the decaying but nonzero scalar field means that $\tilde{\mu}(r)$ is not constant in any region. If one does not assume GR, the changing values of $\tilde{\mu}(r)$ and $\tilde{\nu}(r)$ have to be considered everywhere, and the optical data should be reanalyzed accordingly. This is quite an involved calculation in itself as the observables on Earth depend on the behavior of $\tilde{\mu}(r)$ everywhere outside the star, and for spontaneous scalarization theories, it has been so far applied to real optical data only in limited sample cases~\cite{Silva:2018yxz,Hu:2021tyw}. 

Despite the subtleties we mentioned above, we will compare our numerically computed values of mass and radius directly to those of Ref.~\cite{Ozel:2015fia}, where GR was assumed. Before we consider the error introduced by this approximation, an immediate issue is which value to use as the measured mass of our neutron star. Unless explicitly stated otherwise, we will primarily use the "surface mass" defined as
\begin{align}\label{eq:surface_mass}
M \equiv \tilde{\mu}(r_s) \ 
\end{align}
to be the quantity that is reported as the neutron star mass in Ref.~\cite{Ozel:2015fia}. The ``radius'' of the neutron star is given by
\begin{align}\label{eq:surface_radius}
R \equiv \tilde{r}_s = A(\phi(r_s)) r_s \ ,
\end{align}
$r_s$ being the radius of the star in the frame of the metric $g_{\mu\nu}$ in ansatz~\eqref{metric_ansatz}, i.e.
\begin{align}
\pt(r_s) = 0 \ .
\end{align}

This choice of mass is sensible as long as $\tilde{\mu}(r)$ does not change significantly outside the star, hence the outside spacetime of a scalarized star is also approximately described by the Schwarzschild metric. This is valid for large parts of the parameter space, but not everywhere. We will consider a second definition of neutron star mass to address this issue, the Arnowitt-Deser-Misner (ADM) mass
\begin{align}\label{eq:adm_mass}
M_{\rm ADM} = \tilde{\mu}(\infty) = \mu(\infty) 
\end{align}
as well. We will repeat some of our analysis by using $M_{\rm ADM}$ as the measured neutron star mass, and compare the results to the standard calculations with $M$ in Eq.~\eqref{eq:surface_mass}. Note that outside the star
\begin{align}\label{eq:mass_range}
M< \tilde{\mu}(r) <M_{\rm ADM}\ ,
\end{align}
hence, the surface and ADM masses are the two extremes if one wants to use a value of $\tilde{\mu}(r)$ at a given radius as an effective mass. Consequently, comparing the results from our two mass definitions will provide a useful estimate for the error arising from our approximation. 

Our main aim in this paper is a preliminary analysis of the effectiveness of mass-radius data in constraining spontaneous scalarization phenomenon in neutron stars. Different choices of prior distributions already provide different results as is the nature of Bayesian analysis, and the uncertainty in our posterior distributions will further increase due to the aforementioned approximations about the spacetime outside scalarized stars. This means any bound we obtain will not be precise, but, we will see that the main physical aspects of our findings are robust for different priors and approximation schemes.

\subsection{Likelihood computation}
\label{sec:bayes}
Our approach very closely follows Ref.~\cite{Ozel:2015fia} which computes likelihoods for a parametrized set of EOS using neutron star mass-radius data. Very similarly, we compute the likelihood for the spontaneous scalarization theory parameters while also including the effects of EOS. We use the mass-radius data for the 14 neutron stars in Ref.~\cite{Ozel:2015fia}, and 2 in Ref.~\cite{Bogdanov:2016nle}. The dataset was downloaded from Ref.~\cite{mass_radius_website}.

In summary, we use a Bayesian inference approach to convert
the data from mass-radius measurements of $N=16$ neutron stars to posterior likelihoods for the parameters 
\begin{align}
\{ \pi \} \equiv (\beta,m_{\phi}, \lambda)\ ,
\end{align}
where $\lambda$ parametrizes the EOS. For each star, the mass-radius data is a likelihood distribution $P_{i}(M,R)$ on the mass-radius plane, where $i=1,\dots,N$. We use Bayes' theorem
\begin{align}
P(\{ \pi \}|\text{data})=\mathcal{N}P(\text{data}|\{ \pi \}){\rm Prior}(\{ \pi \}),
\end{align}
where $\mathcal{N}$ is a normalization constant and ${\rm Prior}(\{ \pi \})$ is the prior likelihood of the parameters. The likelihood of the joint data can be written as a product of the likelihoods of the data from each individual star:
\begin{align}
P(\text{data}|\{ \pi \})=\prod_{i=1}^{N}P(\text{data}_{i}|\{ \pi \}).
\end{align}
Here, $\text{data}_{i}$ stands for a single neutron star mass-radius measurement $P_{i}(M,R)$. 

Neutron stars on the mass-radius curve (Fig.~\ref{mr_curve_beta_low}) typically form a one-parameter family for the central density $\rt_c$, which can be used as an auxiliary quantity. That is, to obtain $\text{data}_{i}$ from the parameters $\{ \pi \}$, we specify and marginalize over $\rt_c$:
\begin{align}
P(\text{data}_{i}|\{ \pi \}) =\mathcal{N}_{1}\int_{0}^{\infty}P(\text{data}_{i}|\{ \pi \},\rt_c){\rm Prior}(\rt_c)d\rt_c.
\end{align}
Since there is a one-to-one correspondence between $\rt_c$ and the neutron star mass $M$ (see Appendix~\ref{sec:app_mr_curve} for some details), and since the radius $R$ can be calculated from $M$ once the parameters $\{ \pi \}$ of the theory are given, the above integral can be turned into an integral over $M$:
\begin{align}
\label{eq: distr of a star, given mass and params}
P(\text{data}_{i}|\{ \pi\}) =\mathcal{N}_{2}\int_{M_{\min}}^{M_{\max}}P_{i}(M,&R(M;\{ \pi \}))  \\
& \quad \times {\rm Prior}(M)dM. \nonumber
\end{align}
The prior likelihood over the mass of each neutron star, ${\rm Prior}(M)$, is flat as in Ref.~\cite{Ozel:2015fia}.

We use three different piecewise polytropic EOS introduced in Ref.~\cite{2009PhRvD..79l4033R}: HB, B, 2B.\footnote{There are also stiffer H EOS and 2H EOS, which we did not use mainly because the other three were sufficient to show the dependence of our results on the EOS, and to manage the computational cost. 2B EOS is ruled out assuming GR since the maximum neutron star mass it theoretically provides is considerably lower than the most massive neutron star observed~\cite{NANOGrav:2019jur}, but spontaneous scalarization raises the maximum mass for this EOS well above the observational value, making it viable again.}  
These demonstrate nuclear matter behavior from the very soft (2B EOS) to the stiffer (HB EOS), and cover a wide region of the known physical possibilities. Using only three EOS is a coarse discretization of the $\lambda$ parameter, but our results are sufficient to observe the general trends for different EOS.

We obtain an individual mass-radius curve by constructing $\sim500$ stars with various radius values, and then repeat this procedure for many different points on the $(\beta, \mphi)$ plane, as well as different EOS. Overall, we computed $\sim 5\times 10^6$ individual scalarized neutron star solutions for our calculations. On the $(\beta,\mphi)$ plane, we covered the values $-100<\beta<-4$ with increments of $2$, and $0<\mphi \lesssim 1.5\times10^{-10}$~eV with increments of $8\times10^{-13}$. Note that higher $|\beta|$ and $\mphi$ values are in the GR-equivalent region, and do not have to be computed repeatedly. This meant we had $\sim 6000$ points on the $(\beta,\mphi)$ plane for each EOS, slight variations being due to changing $m_c(\beta)$. 

Such a large number of computations can become prohibitive if efficient methods are not employed. We explain the details of our numerical approach in the appendices. The $(\beta,\mphi)$ parameter space has not been explored before this study for the most part, thus we used a relatively fine grid to make sure that we do not overlook any major features. However, future studies might utilize a more efficient grid structure, and decrease the number of points on the $(\beta,\mphi)$ plane potentially by an order of magnitude or more.

\subsection{Choices of prior distribution}
\label{sec:prior}
There are very few known constraints on the parameters of massive spontaneous scalarization such as those we will discuss in Sec.~\ref{sec:mphi_constraint}. This means, essentially all positive values of $-\beta$ and $\mphi$ are apriori viable choices.\footnote{Spontaneous scalarization can occur for $\beta>0$, but the resulting stars are known to be unstable for our choice of $A(\phi)$~\cite{Mendes:2014ufa,Mendes:2016fby}, hence we will not consider this part of the parameter space. We are interested in lower bounds on $\beta$, which means adding $\beta>0$ values to the consideration would only make these lower bounds stronger. Binary observations that constrain the $\mphi=0$ case~\cite{2013Sci...340..448A,Freire:2012mg} also rule out very low mass values $\meff \lesssim 10^{-16}$~eV~\cite{Ramazanoglu:2016kul}, hence this part of the parameter space is also largely ruled out. However, this extremely low bound will not play an important role in our discussion.} This is radically different from the case of constraining the EOS where one has restrictions on the theory parameters from other observations, nuclear theory and fundamental physics~\cite{Ozel:2015fia}. This makes the choice of the prior probability distribution nontrivial.

Unless explicitly mentioned otherwise, we will use the following prior distribution on the $(\beta, \mphi, \lambda)$ parameter space
\begin{align}
{\rm Prior}^{\rm MC}(\beta, \mphi, \lambda) &= \mathcal{N}\ \frac{e^{-\mphi^2/(2\mu^2)}}{\sqrt{\pi\mu^2/2}}\nonumber\\
&\times \Theta(\beta_\text{max} - \beta) \Theta(\beta-\beta_\text{min}) \,
\label{eq:prior_mc}
\end{align}
where $\Theta$ is the step function, $\beta_\text{min}=-100$, $\beta_\text{max}=0$ and $\mu = 10^{-10}$~eV. We shall call this distribution the ``mass cutoff prior.'' $\beta$ has a flat distribution in the interval $[\beta_\text{min}, \beta_\text{max}]$, and all EOS are also equally likely since there is no dependence on $\lambda$. $\beta$ is a dimensionless quantity, which would naturally be expected to be order-of-unity, hence $\beta_\text{min}=-100$ is a sufficiently large value of theoretical interest. We will further discuss the effects of using a finite interval on our efforts to constrain $\beta$ in the results. As for $\mphi$, we have a half-normal distribution, where $\mu$ acts as a soft cutoff, e.g. values $\mphi \gg \mu$ are considered apriori unlikely. 

The choice of the mass cutoff in our prior needs further elaboration since our lack of clear criteria in choosing the prior might seem to suggest a flat distribution on $\mphi$ as well. However, recall that static neutron star structures, hence also the mass-radius diagrams in our scalar-tensor theory, are identical to those of GR if $\mphi>m_c(\beta)$ for any given $\beta$ where $m_c \sim \sqrt{|\beta|}$ asymptotically. This means, for each EOS, an infinite region of the $(\beta,\mphi)$ parameter plane will have the same evidence in our Bayesian analysis, which will be the Bayesian evidence for GR (see Fig.~\ref{m_c_plot}). Therefore, these points cannot be distinguished from each other, or from GR. More importantly, this GR-equivalent region is unbounded, and would completely dominate the posterior probability distribution if we use a flat or any other non-normalizable prior on $\mphi$, such as a log-flat prior. Therefore, if we marginalize our posterior likelihood over $\mphi$, all $\beta$ would be more or less equally likely, which necessitates some form of cutoff value for $\mphi$ to avoid this triviality.

What is the basis of our cutoff choice of $\mu = 10^{-10}$~eV? Note that in the limit of very large $\mu$, we go back to the trivial case of GR domination in the posterior. Even if we have a mass cutoff with a normalizable prior distribution, the GR-equivalent region would still dominate and wash off any information from the mass-radius data if $\mu$ is large enough. For the $\beta \in [-100,\ 0]$ interval we use, the GR region would dominate for the choice of $\mu=10^{-9}$~eV as we will see in our results. In short, if we admit that any $\mphi \lesssim 10^{-9}$~eV value is more or less equally valid, there is no need for the Bayesian analysis, and one cannot constrain spontaneous scalarization, specifically $\beta$.

On the other hand, the limit of very small $\mu$ would result in a posterior distribution identical to the case of massless scalars, $\mphi=0$. This is due to the fact that scalarized neutron star structures are nearly identical for massless scalars and massive scalars of very low mass, $\mphi \lesssim 10^{-14}$~eV. 

In summary, the behavior of the $\mu \to \infty$ limit is known without any Bayesian analysis, and the $\mu \to 0$ limit can be obtained by the simpler analysis of only considering $\beta$ as a parameter with $\mphi=0$. Hence, we chose $\mu = 10^{-10}$~eV as an intermediate cutoff value in order to understand if we can constrain massive scalarization while avoiding the trivial results of $\mu \to \infty$. We will separately analyze the $\mphi=0$ case in the following section, which is also going to provide the behavior of the $\mu \to 0$ limit. We will also demonstrate this explicitly by considering a lower $\mu$ value as well, but unless stated otherwise, all our results are for $\mu = 10^{-10}$~eV.

A second approach to address the above problem originating from the GR-equivalent region is ignoring this part of the parameter space altogether. Our modified theory is reduced to GR in terms of the neutron star mass-radius relationship when $\mphi>m_c(\beta)$. This means an alternative theory to GR uses two additional parameters without bringing in any explanatory power. We can ``manually'' apply Occam's razor by assigning such cases zero prior probability. By the same logic, we can also ignore $\beta>\beta_c$ for which there is no scalarization either, and obtain what we call ``Occam's prior''
\begin{align}
{\rm Prior}^{\rm Occ}(\beta, \mphi, \lambda) &= \mathcal{N}\
\frac{\Theta(m_c(\beta, \lambda)-\mphi)}{m_c(\beta, \lambda)} \nonumber\\ &\times \Theta(\beta_c(\lambda)-\beta) \Theta(\beta - \beta_\text{min})\ .
\label{eq:prior_occam}
\end{align}
Here, for any given $\beta$ value, $\mphi$ has a flat distribution in the interval $\mphi \in [0, m_c(\beta)]$, and any theory parameter where scalarization does not occur is disregarded.\footnote{Note that the theory we are studying is identical to GR for our dataset of neutron star mass-radius data when $\mphi>m_c(\beta)$, but the theory itself does not reduce to GR completely as we discussed before. There can be, in principle, other observables that can distinguish the two theories. However, the defining aspect of the theory, scalarization, is lost in this part of the parameter space, hence ignoring it in Occam's prior has a philosophical motivation.} This prior can be thought of as a hard cutoff for $\mphi$ as opposed to the soft cutoff of the mass cutoff prior in Eq.~\eqref{eq:prior_mc}, together with a philosophical basis for the choice of the cutoff value, based on Occam's razor. The prior is chosen such that $\beta$ also has a flat distribution if we marginalize over $\mphi$.

As we mentioned before, we mainly present the results for the mass cutoff prior in Eq.~\eqref{eq:prior_mc}. We also report the differences for the other prior choices, and compare them for some select cases. The outcomes of our Bayesian analysis are qualitatively similar for all priors as will be apparent in the coming section. We will investigate the massless scalar case $\mphi=0$ separately, where our two prior distributions are simply the same ones where the dependence on $\mphi$ is disregarded.

\section{Results}\label{results}
\subsection{Massless scalar fields}
\label{results_massless}
We start with applying our methodology to spontaneous scalarization with massless scalars ($\mphi=0$), which is parametrized by $\beta$ in Eq. \eqref{eq:conformal_function}, and $\lambda$, which stands for the EOS. After we perform the Bayesian analysis of the previous section using the data, we calculate the marginal posterior distribution over $\beta$, and also consider the constant $\lambda$ slices to better understand the effect of EOS on the results:
\begin{align}
P_{\rm marg}(\beta) &\equiv \sum_{\lambda}P(\beta,\lambda|{\rm data}) \nonumber \\
p^\lambda_{\rm cond}(\beta)
&\equiv  P(\beta,\lambda|{\rm data})
\label{eq:prob_massless}
\end{align}
Note that $p^\lambda_{\rm cond}$ can also be seen as an unnormalized conditional posterior distribution of $\beta$ for a given $\lambda$. These can be seen in Fig.~\ref{beta_marginal_massless}. $P_{\rm marg}(\beta)$ with the mass cutoff prior for $\mphi=0$, i.e. the flat prior on $\beta \in [-100,\ 0]$, provides the bound
\begin{align}
\beta > -9.8  \quad (\mphi=0)
\label{eq:beta_bound_massless}
\end{align}
at $95\%$ confidence. When we restrict Occam's prior to $\mphi=0$ (flat $\beta$ on $[-100,\ \beta_c(\lambda)]$), the same bound is $\beta>-11.6$, and if we use a log-flat prior on $\beta \in [-100,\ \beta_c(\lambda)]$) it becomes $\beta>-11.2$. This is the first major result of our study, namely, mass-radius data can be used to constrain spontaneous scalarization parameters in the case of massless scalars.
\begin{figure}
\includegraphics[width=.5\textwidth]{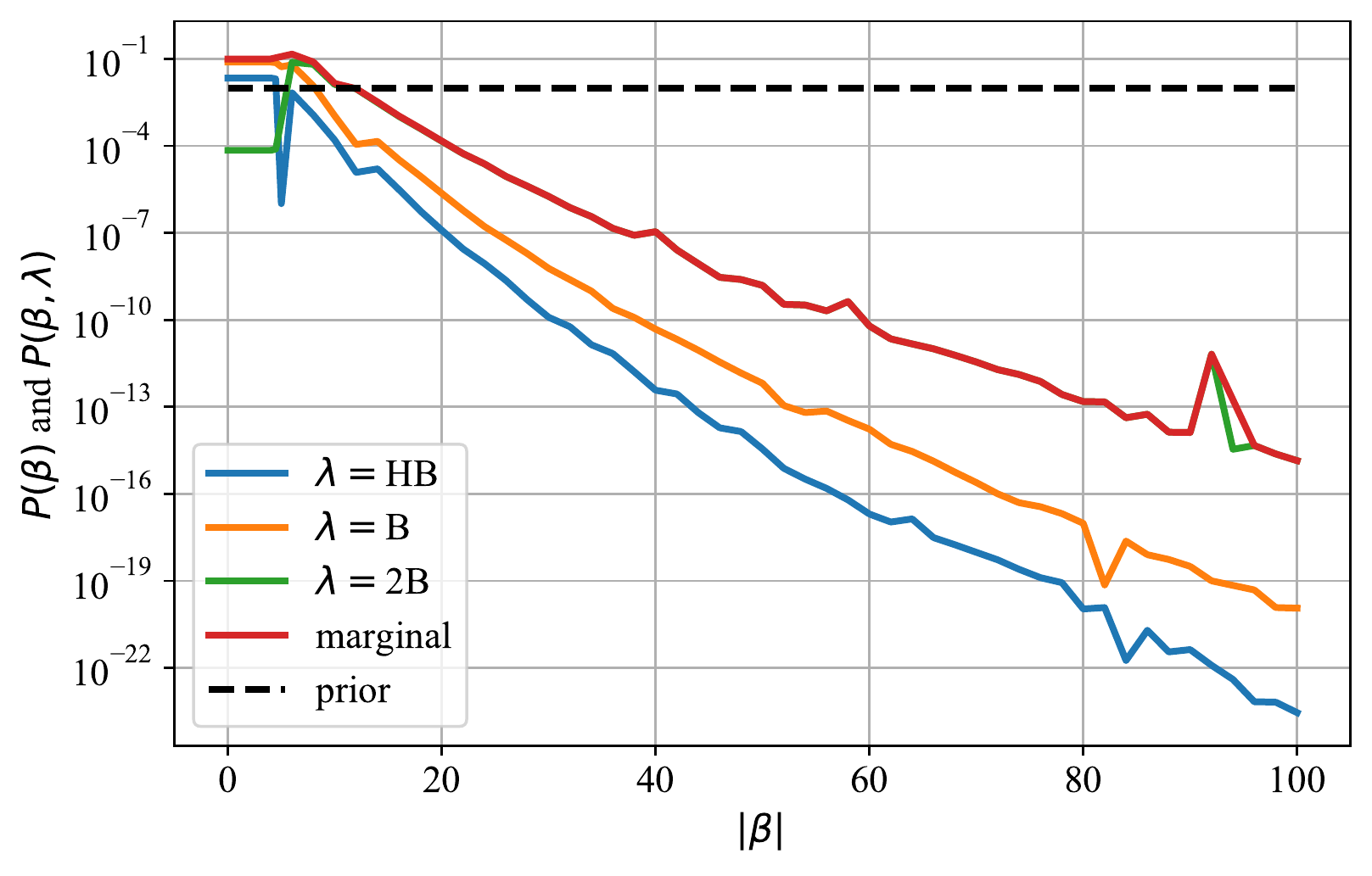}
\caption{
Conditional posterior probability density $p^\lambda_{\rm cond}(\beta)$ for each EOS and the marginalized posterior probability density $P_{\rm marg}(\beta)$ (Eq.~\eqref{eq:prob_massless}) for a massless scalar field ($\mphi=0$) and the mass cutoff prior. The mass-radius data can effectively constrain $\beta$ whether all EOS are considered together or each EOS is taken separately. Cumulative probabilities for each EOS can be seen in Table.~\ref{tableEOS}. Our numerical error leads to some small spikes at high $\beta$, which do not affect our bounds (also see Appendix~\ref{sec:interpolation}).
}
\label{beta_marginal_massless}
\end{figure}

The mass-radius data effectively constrains each EOS we considered, and we expect our results to be valid for more general classes of EOS as well. There is a slight trend in Fig.~\ref{beta_marginal_massless} where stiffer EOS, HB being the stiffest, are more easily constrained compared to the softer ones, 2B being the softest. Evaluating the likelihood of EOS is not a major aim of this work, and its inclusion in our analysis is mainly to ensure that our results are valid independent of this unknown parameter. However, we tabulate the marginal probabilities for each EOS in Table.~\ref{tableEOS} for the sake of completeness.
\begin{table}
\centering
\begin{tabular}{|c || c | c | c | c| } 
 \hline
 EOS($\lambda$) & GR & $\mphi=0$ (flat) & mass cutoff & Occam's \\ [0.5ex] 
 \hline\hline
HB & 0.22 & 0.04 & 0.22 & 0.22 \\
 B & 0.78 & 0.60 & 0.55 & 0.63 \\
2B & $<10^{-3}$ & 0.36 & 0.23 & 0.15  \\ [1ex]
 \hline
\end{tabular}
\caption{Marginal probabilities for each EOS ($P_{\rm marg}(\lambda)  \equiv \int_0^\infty d\beta P(\beta,\lambda|{\rm data})$, and the equivalent version for the massive scalar) for GR, spontaneous scalarization without mass (flat prior on $\beta \in [-100,\ 0]$), and spontaneous scalarization with mass for the two prior choices described in Eqs.~\eqref{eq:prior_mc} and~\eqref{eq:prior_occam}. We only consider a few EOS, mainly to assess their effect on our results for alternative theory parameters, hence these numbers do not provide strong conclusion for the relative likelihoods of the EOS.}
\label{tableEOS}
\end{table}

\begin{figure}
    \includegraphics[width=0.49\textwidth]{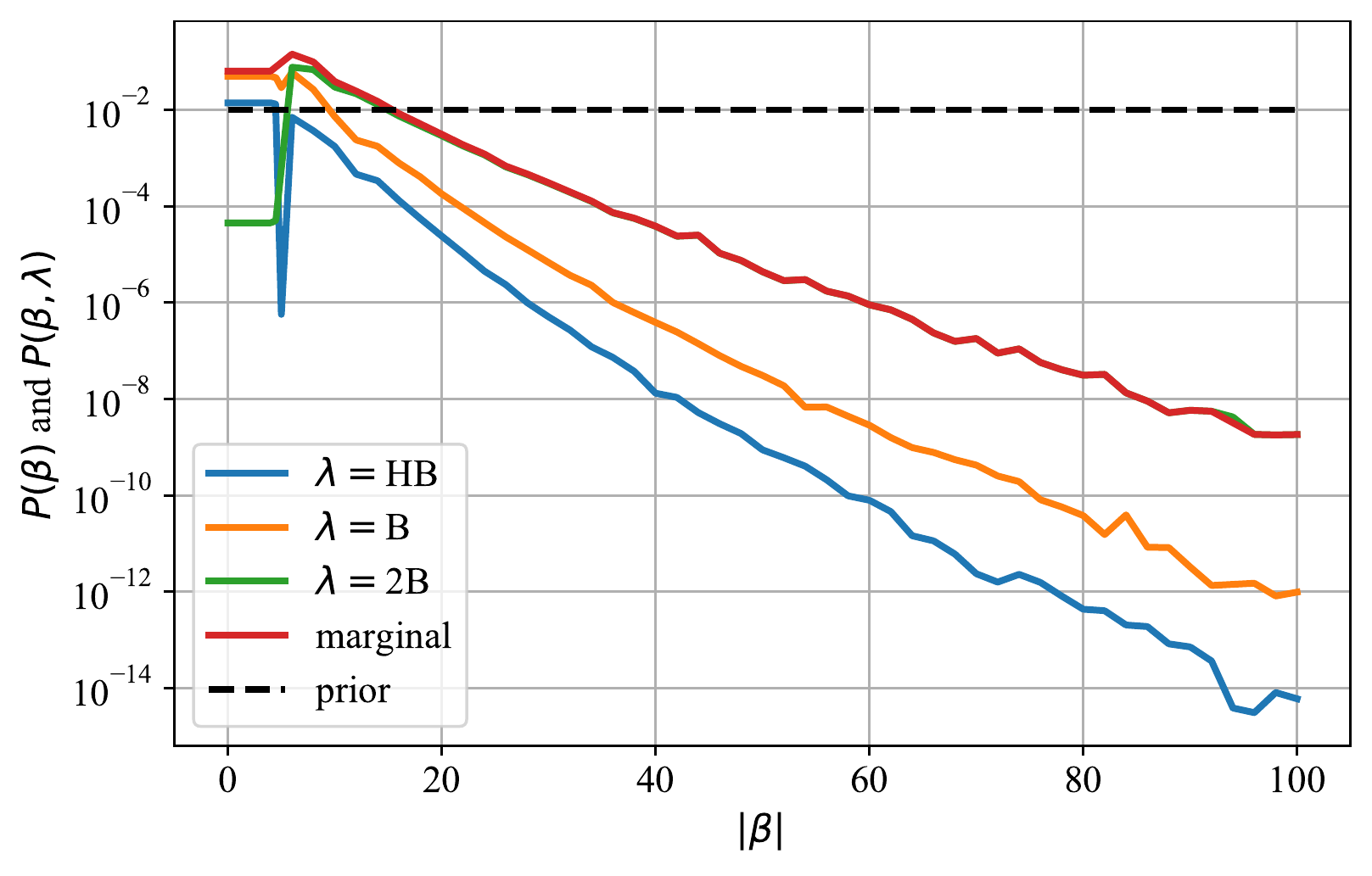}
    \caption{
    Same as Fig.~\ref{beta_marginal_massless}, but ADM mass in Eq.~\eqref{eq:adm_mass} is used as the measured neutron star mass. The qualitative behavior is the same as before, but a different bound of $\beta > -14.7$ is obtained (compare to Eq.~\eqref{eq:beta_bound_massless}).}
    \label{fig:ADM_marginal_massless}
\end{figure}
We also repeated the Bayesian analysis by using the ADM mass in Eq.~\eqref{eq:adm_mass} as the measured neutron star mass. Results can be seen in Fig.~\ref{fig:ADM_marginal_massless}, which provide the $95\%$ confidence lower bound
\begin{align}
    \beta > -14.7, \quad (\mphi = 0,\ \textrm{ADM mass})\ .
    \label{eq:beta_bound_massless_ADM}
\end{align}
This bound is the same order of magnitude with that of Eq.~\eqref{eq:beta_bound_massless} which uses the surface mass in Eq.~\eqref{eq:surface_mass}, and the qualitative behavior of the posterior distribution in Fig.~\ref{fig:ADM_marginal_massless} is similar as well. Nevertheless, the bounds in Eq.~\eqref{eq:beta_bound_massless_ADM} and~\eqref{eq:beta_bound_massless} differ by a factor of order-of-unity. This suggests that our approximations for the neutron star mass can only provide crude values, and more detailed computations that relate the optical signals to the components of the metric in the scalar-tensor theory, such as those in Refs.~\cite{Silva:2018yxz,Hu:2021tyw}, are likely needed to obtain more precise bounds as we explained in Sec.~\ref{sec:mass_radius_curves}.

All results considered together, we reach the crude lower bound
\begin{align}
    \beta \gtrsim -15, \quad (\mphi = 0)\ .
    \label{eq:beta_bound_massless_total}
\end{align}
This is much weaker than $\beta>-4.35$ obtained by binary star observations~\cite{2013Sci...340..448A,Freire:2012mg}\footnote{More recent work claims to rule out spontaneous scalarization completely for massless scalars~\cite{Zhao:2022vig}.}. However, we clearly demonstrate the power of mass-radius data in constraining scalar-tensor theories. Our current analysis uses only 16 neutron stars, and our bounds can improve as this number is expected to increase in the near future~\cite{Bogdanov:2019ixe,Bogdanov:2019qjb,Bogdanov:2021yip}. Also note that the constraint in Eq.~\eqref{eq:beta_bound_massless_total} is well below the cutoff value of $\beta_\text{min}=-100$ in our prior, and together with the fact that the posterior distribution dies off exponentially with growing $|\beta|$, this means $\beta_\text{min}$ is sufficiently low for this analysis.

\subsection{Massive scalar fields}
\label{results_massive}
%
\begin{figure}
    \centering
    \includegraphics[width=0.49\textwidth]{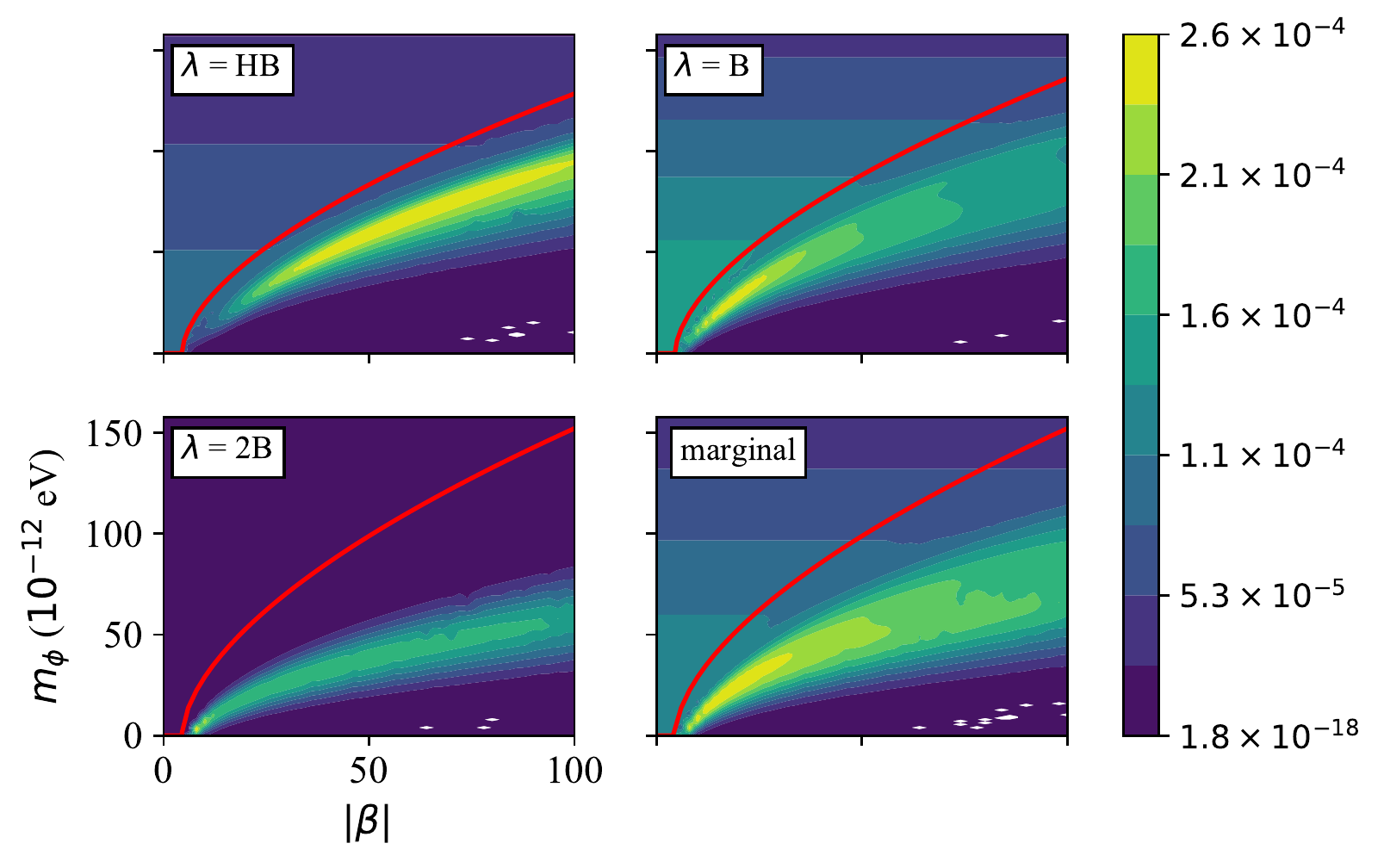}
    \includegraphics[width=0.49\textwidth]{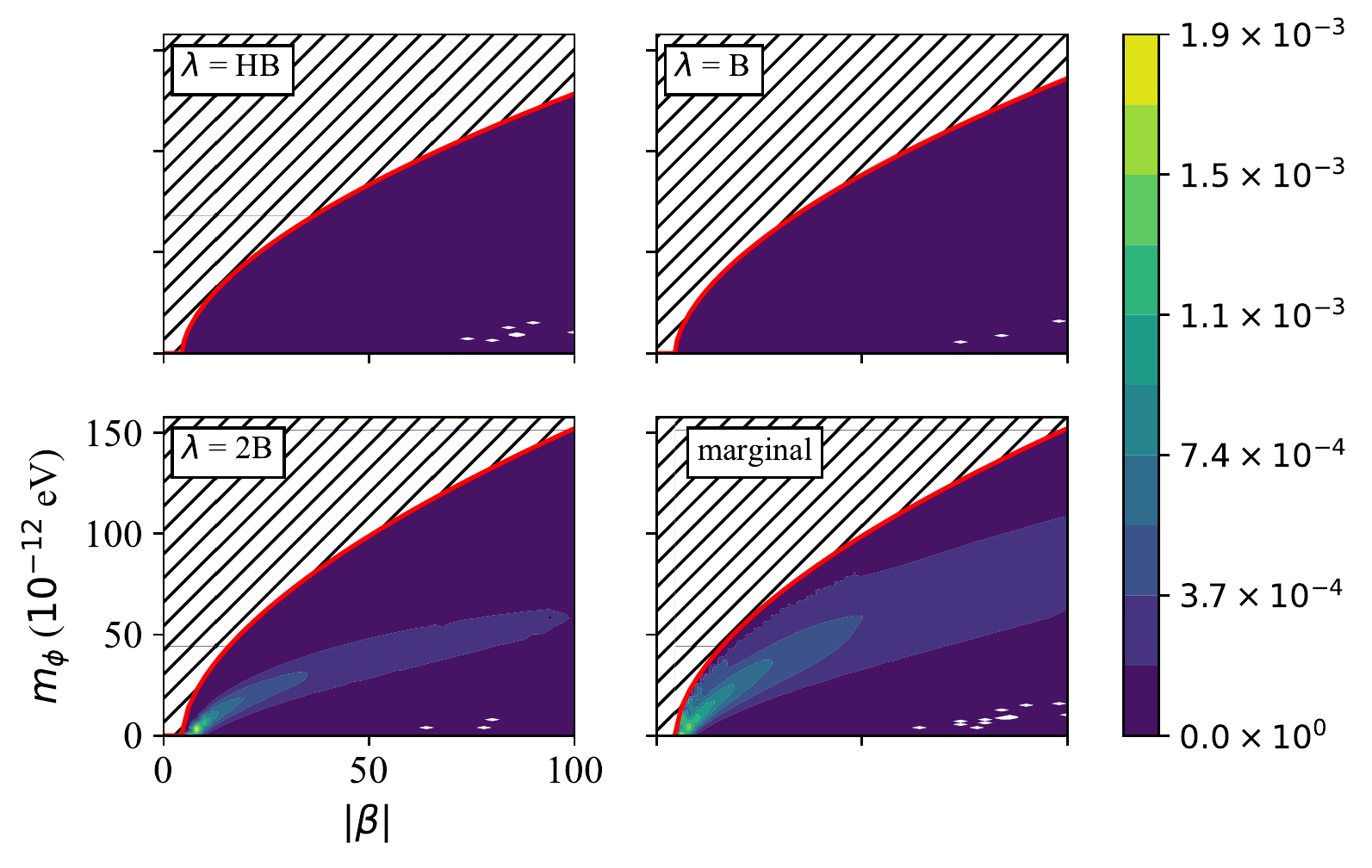}
    \caption{
    Conditional distributions $p^\lambda_{\rm cond}(\beta,\mphi)$ and the marginalized  distribution $P_{\rm marg}(\beta,\mphi)$ (Eq.~\eqref{eq:marginal_beta_mphi}) for the mass cutoff prior (upper) and Occam's prior (lower). The high likelihood region below the red critical mass curves $m_c(\beta,\lambda)$ is present for all EOS and priors, even though they are not visible in the HB and B EOS for Occam's prior due to the common colormap. The white spots on the lower right part of the plots are due to the failure of our numerical computation in isolated extremely scalarized cases, which is discussed in Appendix~\ref{sec:interpolation}.
    }
    \label{mphi_beta_posterior_cutoff}
\end{figure}
The massive scalar case can be analyzed similarly to that of the massless one, where $\mphi$ is added as a parameter. The marginal and unnormalized conditional probabilities of $(\beta,\mphi)$ are given by
\begin{align}
P_{\rm marg}(\beta,\mphi) 
&\equiv \sum_{\lambda}P(\beta,\mphi,\lambda|{\rm data}) \nonumber \\
p^\lambda_{\rm cond}(\beta,\mphi)
&\equiv P(\beta,\mphi,\lambda|{\rm data}) \ , 
\label{eq:marginal_beta_mphi}
\end{align}
respectively, just like the massless case in Eq.~\ref{eq:prob_massless}. We can also further marginalize over $\mphi$ to obtain the marginal and unnormalized conditional probabilities of $\beta$ alone
\begin{align}
P_{\rm marg}(\beta) 
&\equiv  \sum_{\lambda} \int_0^\infty d\mphi\ 
P(\beta,\mphi,\lambda|{\rm data}) \nonumber \\
p^\lambda_{\rm cond}(\beta)
&\equiv \int_0^\infty d\mphi\ 
P(\beta,\mphi,\lambda|{\rm data})\ .
\label{eq:marginal_beta_massless}
\end{align}
Defining analogous one-dimensional distributions for $\mphi$ is straightforward, but we will not attempt this. The fact that the modified theory that features spontaneous scalarization is equivalent to GR for high $\mphi$ values makes the choice of priors for $\mphi$ highly nontrivial as we discussed, and in turn, we do not see a simple way to interpret the marginal posterior distribution of $\mphi$, which is strongly dependent on the choice of prior.

The posterior distributions for the parameters are radically altered for massive scalars as can be seen in Fig.~\ref{mphi_beta_posterior_cutoff}. The most important feature of the posterior distribution  is the fact that there is a high-likelihood region on the $(\beta,\mphi)$ parameter space between $\mphi=0$ and $\mphi=m_c(\beta)$. The conditional probability for each EOS ($p^\lambda_{\rm cond}(\beta,\mphi)$) and the marginal probability ($P_{\rm marg}(\beta,\mphi)$) clearly show a peak likelihood that occurs at relatively high values of $|\beta|$. This is the case for both the mass cutoff and Occam's priors in Eqs.~\eqref{eq:prior_mc} and ~\eqref{eq:prior_occam}, respectively.

The high-likelihood region may not look necessarily surprising since we have more parameters in the case of massive scalars, hence more degrees of freedom to better accommodate the observational data. However, the existence of this region is a result of a specific aspect of the theory we are investigating, rather than being a consequence of simply having more parameters. Recall that $\beta$ and $\mphi$ have opposing effects on scalarization: While more negative $\beta$ values facilitate scalarization, we can also increase $\mphi$ for any given $\beta$ to obtain a theory where mass-radius curves are not radically different from that of GR. Our scalar-tensor theory is equivalent to GR on the critical mass curve $m_c(\beta)$ as far as the star structure is concerned. Moving off from the critical mass curve down to the region of scalarization, $(\beta,\mphi)$ can be used to tweak the star structure continuously, and provide a better fit to the data. However, this trend does not continue indefinitely, and the maximum likelihood for any given $\beta$ starts to decrease at the high $|\beta|$ region. 

\begin{figure}
    \centering
    \includegraphics[width=0.49\textwidth]{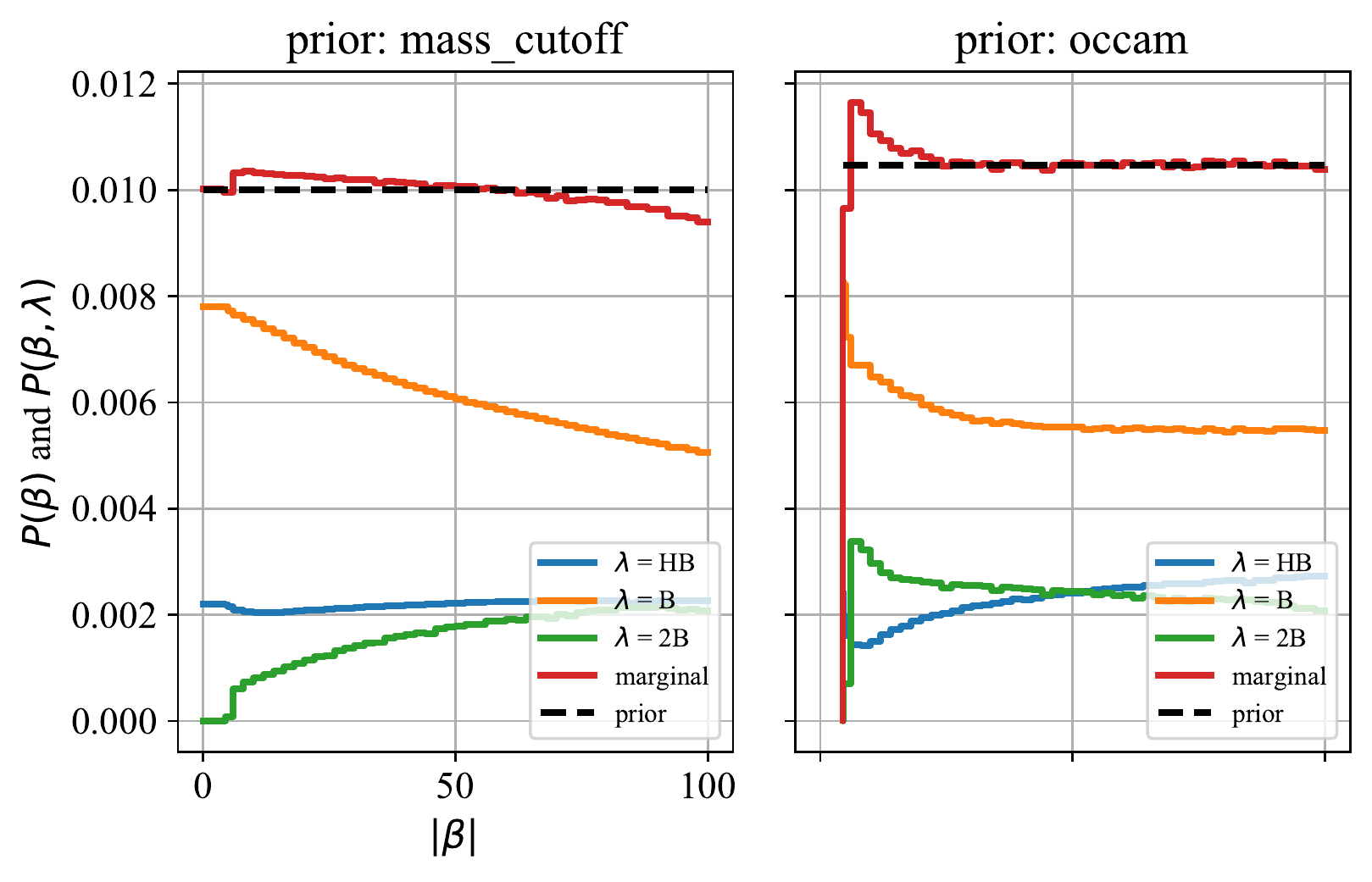}
    \caption{
    The marginal ($P_{\rm marg}(\beta)$) and conditional ($p^\lambda_{\rm cond}(\beta)$) probabilities of $\beta$ alone (Eq.~\eqref{eq:marginal_beta_massless}), obtained by integrating over $\mphi$ in Fig.~\ref{mphi_beta_posterior_cutoff}. Left: Mass cutoff prior. Right: Occam's prior. The marginal likelihood is mostly flat in both cases, which means we cannot put a lower bound on $\beta$ or obtain any other significant information when $\mphi \neq 0$.
    }
    \label{beta_marginal_massive}
\end{figure}
The high-likelihood region increases the marginal probability of $\beta$ for large $|\beta|$ compared to the $\mphi=0$ case, thus the constraints we can obtain are clearly less strict. We quantify this in three ways. First, we marginalize over $\mphi$ (Eq.~\eqref{eq:marginal_beta_massless}), whose results are plotted in Fig.~\eqref{beta_marginal_massive} for both the mass cutoff and Occam's priors. Comparing this with its counterpart for the massless case, Fig.~\eqref{beta_marginal_massless}, shows the radical effect of scalar field mass on the posterior distribution. Namely, it is not possible to put any constraints on $\beta$ based on neutron star mass-radius data independently of $\mphi$. There is a slight preference for lower $|\beta|$ overall, but the deviation from the prior is not significant. This shows that even though the probability density peaks at $\beta \sim -25$ on the $(\beta,\mphi)$ plane (see Fig.~\ref{mphi_beta_posterior_cutoff}), the width of the high-likelihood region behaves in such a way that when we marginalize over $\mphi$, $P_{\rm marg}(\beta)$ stays more or less flat. In principle, we can calculate a $95\%$ confidence lower bound for $\beta$ since we consider only the finite interval $\beta \in [-100,\ 0]$, but there is no tail for $P_{\rm marg}(\beta)$. It is clear that the $\beta<-100$ region which we do not consider is still highly likely, and the bounds on $\beta$ based on the $\beta>-100$ interval would not be meaningful.

The second way to quantify the effect of the high likelihood region is investigating the posterior for $\beta$ when we assume various different values of $\mphi$, that is, repeating our analysis for the massless scalar but this time setting $\mphi$ to nonzero values, one at a time. Results can be seen in Fig.~\ref{mphi_nonzero} for the flat prior on $\beta$. We can explicitly see that $\beta$ can still be constrained in terms of a lower bound when $\mphi \lesssim 3\times10^{-11}$eV, where the high likelihood region is not dominant. However, as higher $\mphi$ are considered, the posterior likelihood first becomes flatter with respect to $\beta$, and even starts to favor higher $|\beta|$ when the high-likelihood region covers this part of the parameter space. Fig.~\ref{beta_marginal_massive} is a weighted sum over the values of $\mphi$, hence combines these effects to provide a mostly flat posterior.

\begin{figure}
    \centering
    \includegraphics[width=0.49\textwidth]{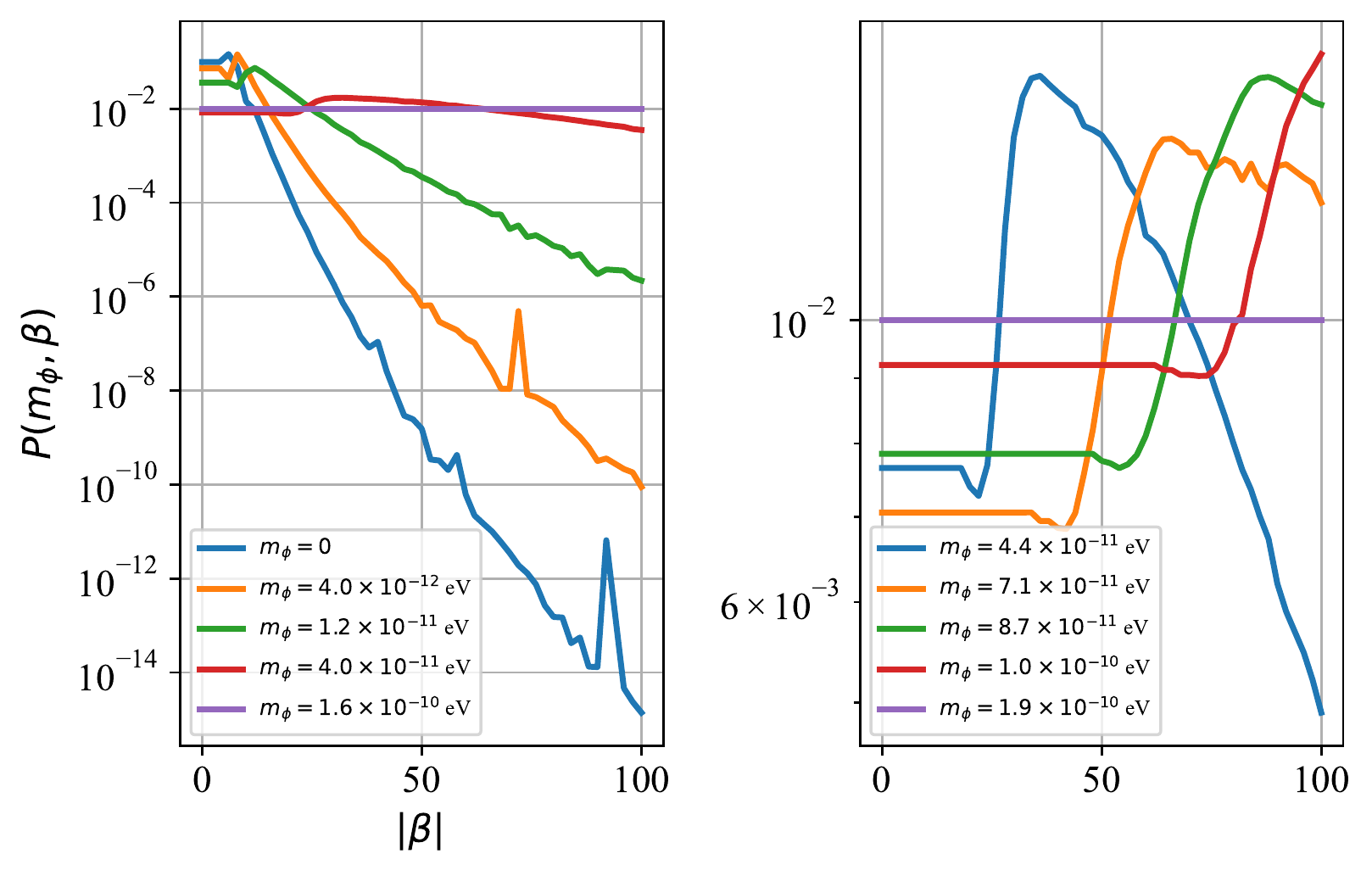}
    \caption{
     The posterior probability density $P(\mphi, \beta)$ (marginalized over all EOS) for various assumed values of $\mphi$ (analogs of the marginal curve in Fig.~\ref{beta_marginal_massless} for different $\mphi \neq 0$). Left: Lower values of $\mphi$ where a lower bound can be obtained for $\beta$, and the single high value of $\mphi =1.6\times10^{-10}$eV as a reference. Right: Higher values of $\mphi$ where the posterior is not particularly informative about $\beta$.  It is clear that one can obtain the bound $\beta \gtrsim -20$ for $\mphi \lesssim 2\times 10^{-11}$eV, but there is no effective bound for higher masses, at least in the interval $\beta \in [-100,\ 0]$. No inference for $\beta$ is possible for $\mphi \gtrsim 10^{-10}$eV due to the GR-equivalent region.
    }
    \label{mphi_nonzero}
\end{figure}
Our computational methods failed more frequently or became extremely slow beyond $\beta \sim -100$, and we could not test this region in detail. Whether an exponential or power law decrease in $P_{\rm marg}(\beta)$ occurs for even more negative values such as $\beta \sim -10^3$ is unknown. If this is the case, such a bound on $\beta$ might look excessively weak, especially when we recall our argument that $\beta$ is a fundamental dimensionless constant of the theory which we would expect to be order-of-unity. However, note that this would be the only known bound on $\beta$ for $\mphi \neq 0$, hence, could be an important first step.
\begin{figure}
    \centering
    \includegraphics[width=0.49\textwidth]{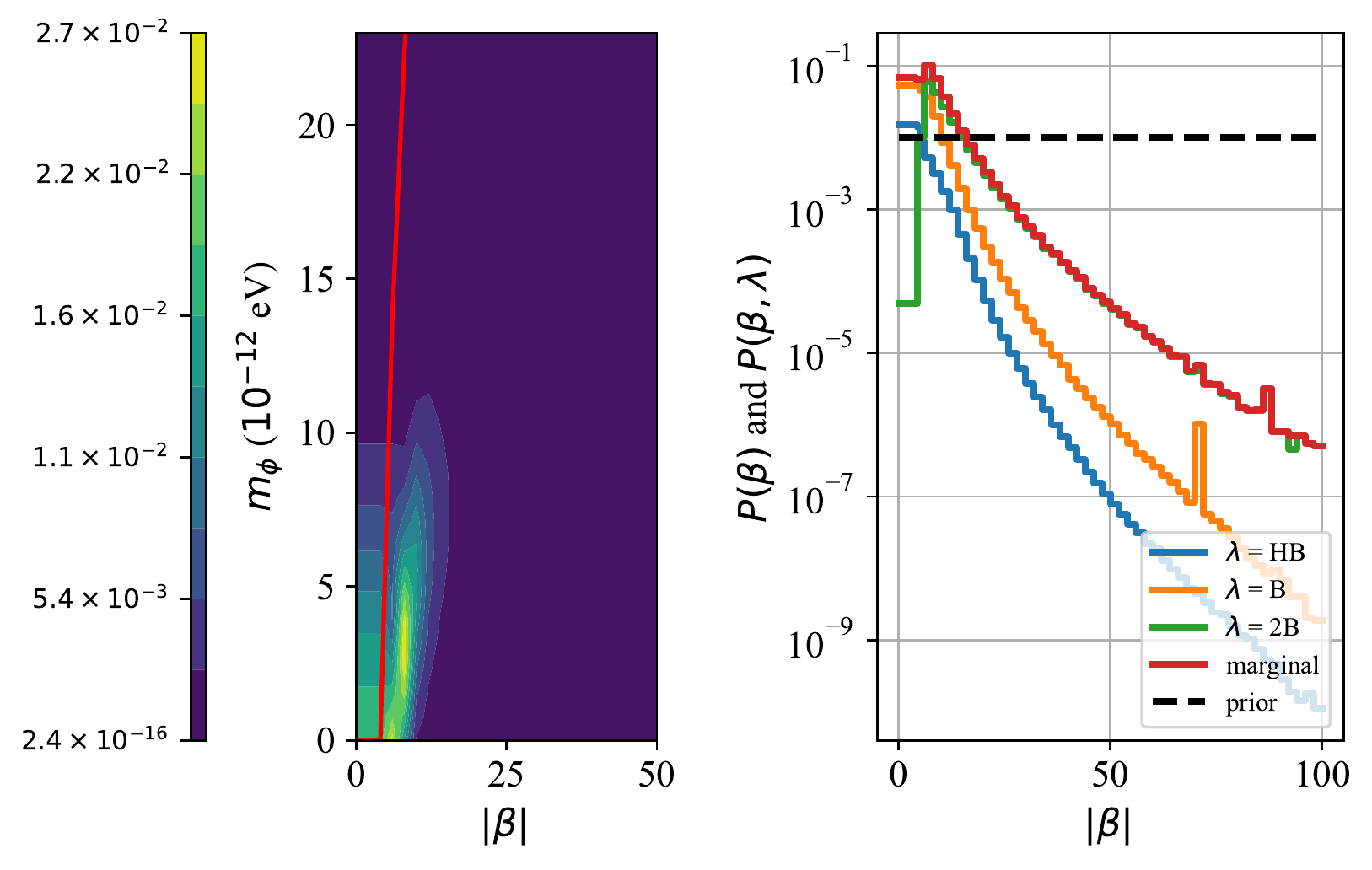}
    \caption{
    Posterior distributions for mass cutoff prior (Eq.~\eqref{eq:prior_mc}) with $\mu = 5 \times 10^{-12}$~eV. Left: $P_{\rm marg}(\beta,\mphi)$ (Eq.~\eqref{eq:marginal_beta_mphi}), analog of Fig.~\ref{mphi_beta_posterior_cutoff}. Right: $P_{\rm marg}(\beta)$ and $p^\lambda_{\rm cond}(\beta)$ (Eq.~\eqref{eq:marginal_beta_massless}), analog of Fig.~\ref{beta_marginal_massive}. Unlike our standard choice $\mu = 10^{-10}$~eV, $\beta$ can be constrained in this case thanks to the exponential decay in the figure on the right. This is expected since the posterior converges to that of the massless scalar (Fig.~\ref{beta_marginal_massless}) in the $\mu \to 0$ limit, where we could put a lower bound on $\beta$.
    }
    \label{marginal_massive_lower_MC_cutoff}
\end{figure}

The fact that mass-radius data cannot constrain spontaneous scalarization independently of the scalar mass in an  unambiguous way is the second major result of this study. One of the culprits of this fact is the $\mu=10^{-10}$~eV value we use as the mass cutoff in the prior. We discussed in Sec.~\ref{sec:prior} that in the $\mu \to 0$ limit, the posterior distribution of $\beta$ converges to that of the massless scalar case where we could put lower bounds on $\beta$ (Eq.~\eqref{eq:beta_bound_massless}). This can be seen explicitly in Fig.~\ref{marginal_massive_lower_MC_cutoff}, where we use the sharp cutoff of $\mu=5 \times 10^{-12}$~eV, and obtain the bound 
\begin{align}
\beta > -17.1  \ \ \ \ (\mphi>0,\ \mu=5 \times 10^{-12}~\textrm{eV})\ .
\label{eq:beta_bound_massive}
\end{align}
with $95\%$ confidence. This is the third way to quantify the effect of the high likelihood region and $\mphi$ in our analysis. 

Despite the bound in Eq.~\eqref{eq:beta_bound_massive}, it is hard to physically justify a specific cutoff $\mu$, hence the sensitivity of our results on this choice is a serious shortcoming. Moreover, Occam's prior has a more natural choice of a mass cutoff in the form of $m_c(\beta, \lambda)$, and it cannot constrain $\beta$ either. Despite these facts, considering lower values of $\mu$ can help in understanding what independent constraints on $\mphi$ can help the mass-radius data to constrain spontaneous scalarization. We will further elaborate on this in Secs.~\ref{sec:mphi_constraint} and~\ref{sec:conclusions}.

The behavior of $p^\lambda_{\rm cond}(\beta)$ shows qualitative variations between the EOS in Fig.~\ref{beta_marginal_massive}, unlike the massless case in Fig.~\ref{beta_marginal_massless}. For example, the likelihood monotonically decreases with increasing $|\beta|$ for B EOS. However, the changes are small at high $|\beta|$ for all EOS such that it is not meaningful to assign a statistical bound on $\beta$ in any case. Nevertheless, dependence on EOS might be another factor to investigate in the future.
\begin{figure}
    \centering
    \includegraphics[width=0.49\textwidth]{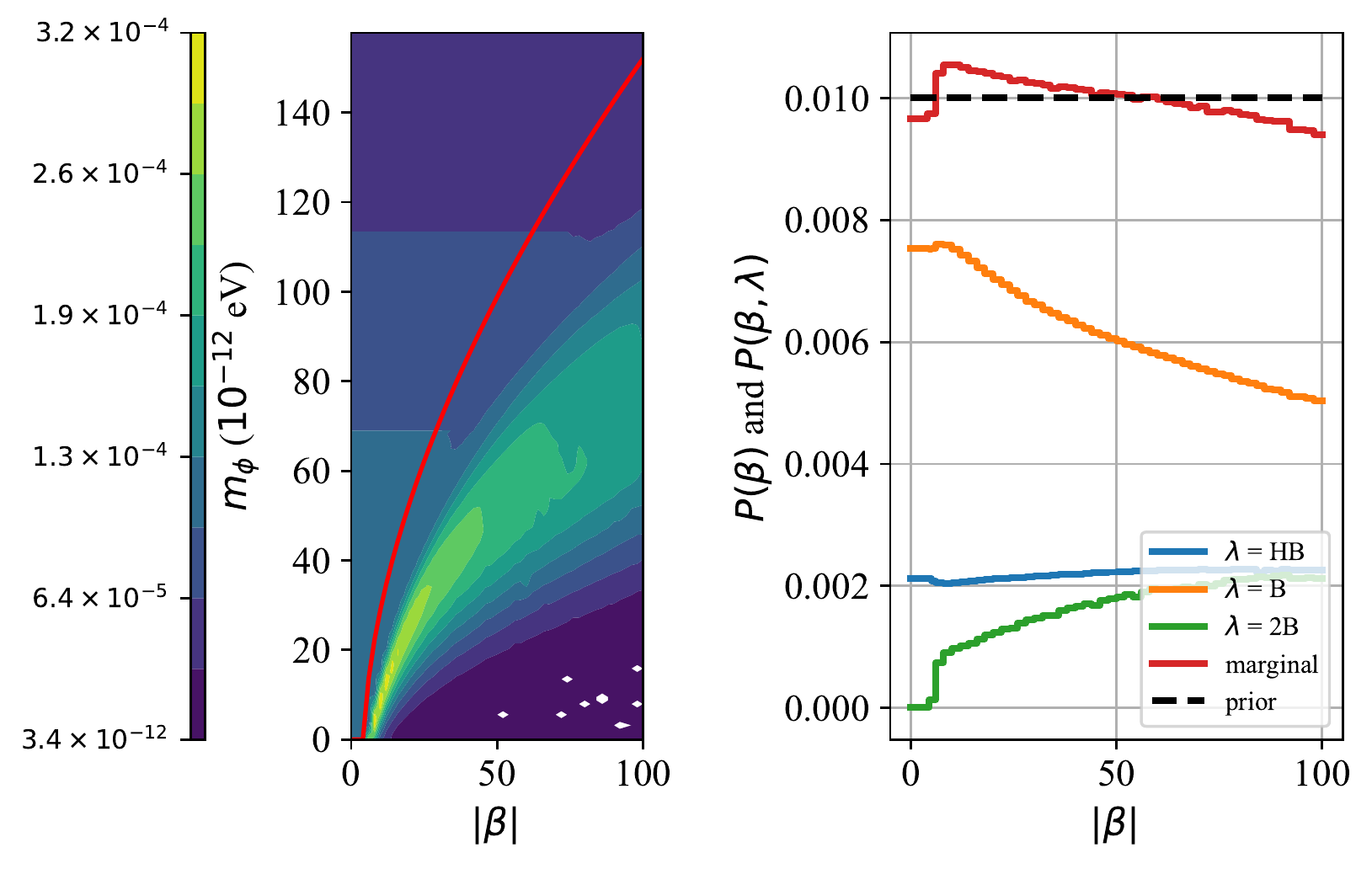}
    \caption{
    Posterior distributions for mass cutoff prior (Eq.~\eqref{eq:prior_mc}) where the ADM mass (Eq.~\eqref{eq:adm_mass}) is used instead of the surface mass (Eq.~\eqref{eq:surface_mass}). Left: $P_{\rm marg}(\beta,\mphi)$ (Eq.~\eqref{eq:marginal_beta_mphi}), the analog of Fig.~\ref{mphi_beta_posterior_cutoff}. Right: $P_{\rm marg}(\beta)$ and $p^\lambda_{\rm cond}(\beta)$ (Eq.~\eqref{eq:marginal_beta_massless}), analog of Fig.~\ref{beta_marginal_massive}. The use of this alternative approximation for the measured neutron star mass has some quantitative differences compared to the case of surface mass, but the general behavior is unchanged. Namely, it is still not possible to put any meaningful constraints on $\beta$ due to a high-likelihood region.
    }
    \label{fig:ADM_marginals_massive}
\end{figure}
Lastly, we repeated our analysis with the ADM mass instead of the surface mass as well. There are differences in the posterior values, but all results are qualitatively similar as can be seen in Fig.~\ref{fig:ADM_marginals_massive}. Overall, this strongly suggests that a more involved analysis of the data, as opposed to our simplifying assumptions about the measured mass in Sec.~\ref{sec:mass_radius_curves}, would still not be able to constrain the theory parameters.

To summarize, the $\beta$ parameter of spontaneous scalarization theories cannot be constrained independently of $\mphi$ using neutron star mass-radius data alone. This is due to the opposite effects of $\beta$ and $\mphi$ on deviations from GR, which lead to a region of high posterior likelihood on the $(\beta,\mphi)$ parameter space. The posterior distribution for $\beta$ does not deviate significantly from the prior as a result, at least within the $0>\beta>-100$ region we investigated. 

\subsection{Effect of additional constraints on $\mphi$}
\label{sec:mphi_constraint}
%
\begin{figure}
\centering
\includegraphics[width=0.49\textwidth]{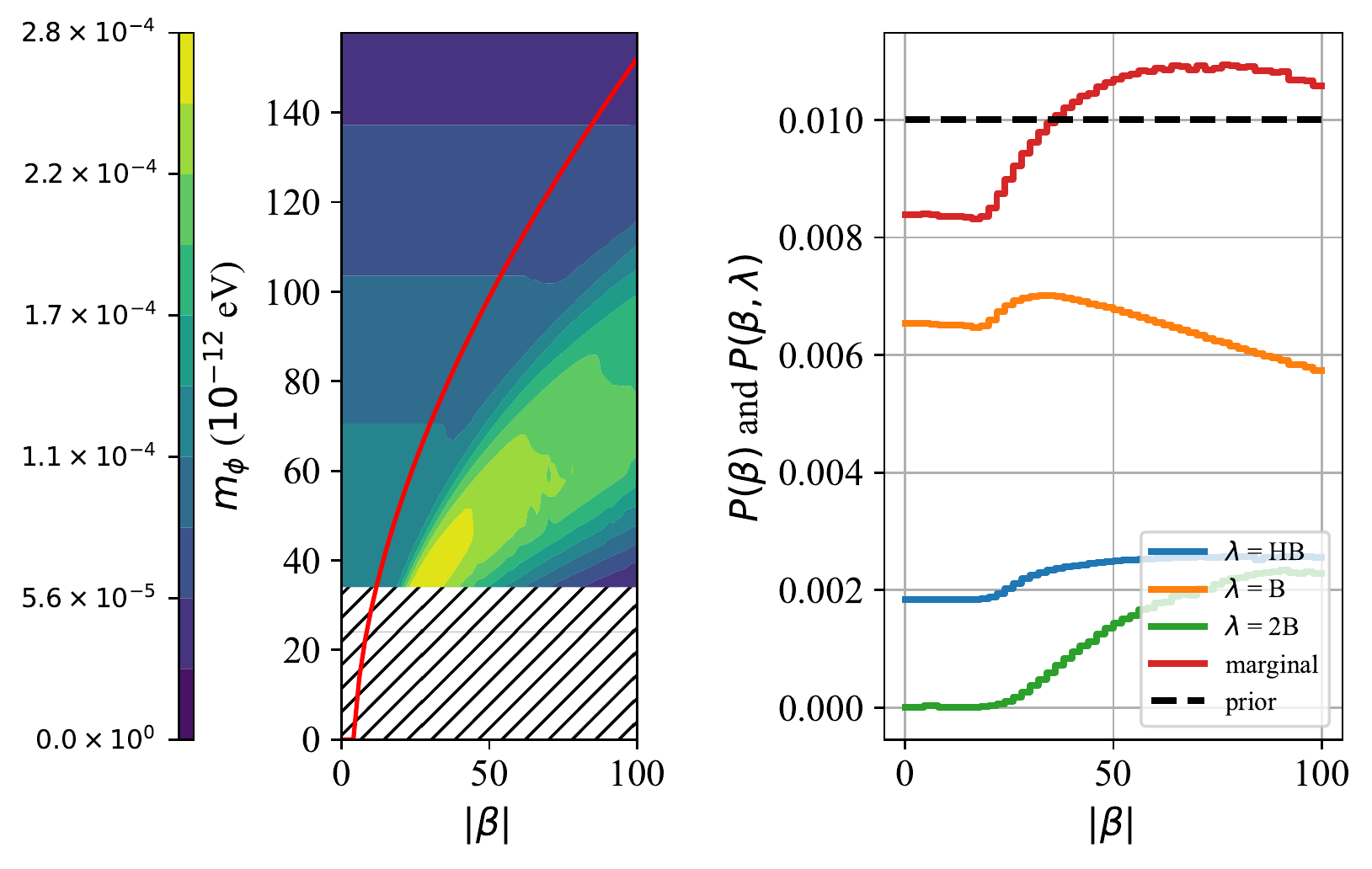}
    \caption{
    Posterior distributions for mass cutoff prior where we also incorporate the additional constraints on $\mphi$ from black hole spin data (traced out region). This is simply obtained by setting the probability distribution to zero in Fig.~\ref{mphi_beta_posterior_cutoff} within the ruled out region and renormalizing. Left: $P_{\rm marg}(\beta,\mphi)$ (Eq.~\eqref{eq:marginal_beta_mphi}), the analog of Fig.~\ref{mphi_beta_posterior_cutoff}. Right: $P_{\rm marg}(\beta)$ and $p^\lambda_{\rm cond}(\beta)$ (Eq.~\eqref{eq:marginal_beta_massless}), analog of Fig.~\ref{beta_marginal_massive}. These additional bounds enhance the deviation of the posterior distribution from the prior, but it is still not possible to obtain meaningful bounds on $\beta$.} 
\label{beta_marginal_mphicons}
\end{figure}
We have exclusively used the neutron star mass-radius data so far, however, additional bounds on $\mphi$ can be obtained from other astrophysical observations. The best bound on $\mphi$ in spontaneous scalarization comes from black hole spin measurements. The tachyonic instability in the action~\eqref{st_action} only occurs in the presence of matter. Thus, the scalar-tensor theory in Eq.~\eqref{st_action} reduces to GR with a minimally coupled scalar field around black holes, which do not carry any stable hair. On the other hand, scalar fields can efficiently grow due to superradiance around spinning black holes, and form unstable, yet long-lived, clouds~\cite{2015CQGra..32m4001B,Brito:2017zvb}. This growth occurs at the expense of the black hole spin, which means that observations of black holes with sufficiently large spins can rule out scalar fields of certain masses. 

The most stringent current bound from black hole spins rule out the scalar mass range
\begin{align}\label{eq:mass_bounds}
3.8\times10^{-14}~{\rm eV} < \mphi < 3.4\times10^{-11}~{\rm eV}
\end{align} 
with $95\%$ confidence~\cite{Stott:2018opm,Stott:2020gjj,Ng:2020ruv}. This is a sizeable part of the $(\beta,\mphi)$ plane where spontaneous scalarization occurs. 

We incorporated these independent constraints by updating our prior likelihood so that it vanishes in the interval in Eq.~\eqref{eq:mass_bounds}. We implement this simply by setting the prior (hence also the posterior) likelihood  within the ruled out $\mphi$ interval to zero, and renormalizing. This is not an exact translation of the data to our analysis since the bounds do not have sharp cutoffs, but the difference is insignificant. 

Even though the ruled out mass interval in Eq.~\eqref{eq:mass_bounds} has a large overlap with the high likelihood region in Fig.~\ref{mphi_beta_posterior_cutoff},\footnote{This is not a coincidence, the relevant mass values for spontaneous
scalarization, $\mphi \lesssim m_c(\beta)$, are at the order of the inverse of the size of a typical neutron star when $\beta$ is order of unity~\cite{Ramazanoglu:2016kul}. Similarly, superradiance is most efficient when the Compton wavelength corresponding to the scalar field mass, which is the inverse of the mass, is comparable to the horizon size of a black hole~\cite{2015CQGra..32m4001B,Brito:2017zvb}. Stellar black hole horizons and neutron stars have similar sizes, hence, the constraints from superradiance are naturally relevant for spontaneous scalarization.} this is still not enough to obtain any bounds on $\beta$ as can be seen in Fig.~\ref{beta_marginal_mphicons}. The additional constraints on $\mphi$ slightly change the posterior distribution of $\beta$ compared to Fig.~\ref{beta_marginal_massive}, but not to the extend that $\beta$ is confined to a particular region of $\beta \in [-100,\ 0]$. There are some features on the posterior distribution of $\beta$ in Fig.~\ref{beta_marginal_mphicons}, which are partly due to our somewhat ad-hoc incorporation of the bounds in Eq.~\eqref{eq:mass_bounds}. One can try to device a new prior distribution based on the ruled out mass range to avoid this, but the main fact that we cannot bound $\beta$ independently of $\mphi$ even with the additional $\mphi$ constraints would not change.

\section{Conclusions}
\label{sec:conclusions}
%
We investigated the possibility of constraining the spontaneous scalarization phenomenon using neutron star mass-radius data. There exists some previous work that uses some aspects of this data such as the maximum allowed neutron star mass~\cite{Sotani:2017pfj,SavasArapoglu:2019eil}, but to our knowledge, this is the first study to utilize all the information in the mass-radius curves for this purpose through Bayesian analysis.

A prerequisite for our statistical computations is obtaining the mass-radius curves for neutron stars over a large portion of the $(\beta,\mphi)$ parameter space and for a variety of EOS. This required numerically solving the TOV equations and computing the structures of $\sim 5\times 10^6$ stars. This is a challenging task, especially at large values of $|\beta|$, and we followed a novel computational approach based on the work of Ref.~\cite{Rosca-Mead:2020bzt}. Statistical part of our work is quite similar to the past attempts to constrain nuclear EOS using the same data~\cite{Ozel:2015fia}. We had to consider the effects of EOS and deviations from GR together, which complicates the analysis. The choice of the prior is also nontrivial for scalarization with massive fields due to the fact that an infinitely large part of the $(\beta,\mphi)$ parameter space provides mass-radius curves that are identical to those of GR.

We showed that neutron star mass-radius data can be used to constrain the $\beta$ parameter of the theory for a massless scalar, and obtained the $95\%$ confidence bound $\beta \gtrsim -15$, which varies around this value depending on the prior and our approximations about the measured neutron star mass. This is weaker than the existing bounds from binary observations~\cite{2013Sci...340..448A,Freire:2012mg,Zhao:2022vig}, but demonstrates a proof of principle for the effectiveness of the mass-radius data in constraining deviations from GR.

Constraining spontaneous scalarization for massive fields is a much more complicated task. We saw that it is not possible to constrain $\beta$ independently of $\mphi$ by solely using the mass-radius data. Aside from not being able to obtain a bound, the posterior distribution of $\beta$ is quite similar to the prior (see Fig.~\eqref{beta_marginal_massive}), hence we cannot infer whether any value of $\beta$ is strongly favored or disfavored. It might be possible that the same analysis provides bounds if it is continued beyond $\beta<-100$, but the significance of such a result is debatable. The dimensionless quantity $\beta$ is typically expected to be order-of-unity, and very large values are apriori less interesting.

Ineffectiveness of the mass-radius data in constraining $\beta$ is surprising, since the $\mphi$ term was partially motivated by avoiding the bounds on spontaneous scalarization from binary observations~\cite{Ramazanoglu:2016kul}, but it seems this term enables the theory to avoid bounds from the mass-radius data as well. Neutron star structures typically deviate radically from the case of GR at even modestly high $|\beta|$ values (see Fig.~\ref{mr_curve_beta_low}), and one would naively expect such large differences, hence such $\beta$ values, to be ruled out, which is not the case. This is an important discovery by itself. We identified the main reason for this surprise to be the opposing nature of the $\beta$ and $\mphi$ parameters on the deviations from GR. Even though the neutron star structure can deviate drastically from the GR case at large $|\beta|$, the deviations can also be suppressed by large values of $\mphi$. Basics of this was known, but we show that the region of the $(\beta,\mphi)$ plane where the two effects largely cancel each other is quite large. This region has relatively high posterior likelihood, which makes ruling out large $|\beta|$ values impossible for high values of $\mphi$. The exception is the case where the prior is mostly confined to $\mphi \lesssim 10^{-11}$~eV. Current independent bounds do not provide such a confinement, but if an independent upper bound on the scalar mass can be obtained by other means, the mass-radius data might become more informative as in the case of $\mphi=0$.

One way to obtain better independent bounds on $\mphi$ is ruling out a larger region of the parameter space using black hole spin observations. In contrast to ruling out high scalar mass values, ruling out $\mphi \lesssim 10^{-10}$~eV could also remove the problematic high-likelihood region in our posterior distribution (see Fig.~\ref{mphi_beta_posterior_cutoff}), which in turn might enable us to better constrain $\beta$. Constraints from black holes are related to the horizon sizes~\cite{2015CQGra..32m4001B,Brito:2017zvb}, and one needs observations of black holes with smaller masses by a factor of a few to increase the current upper bound of the ruled out region, $3.4\times 10^{-11}$~eV (Eq.~\eqref{eq:mass_bounds}), to the desired value. This might be possible, but known astrophysical formation mechanisms and population statistics do not favor such low black hole masses~\cite{LIGOScientific:2021psn}. Future observations will be the ultimate judge on this issue.

Our work can be extended in various directions. The first obvious direction is considering more numerous and precise mass-radius data. There is an organized effort to extend the dataset we considered by using X-ray observations of more neutron stars~\cite{Bogdanov:2019ixe,Bogdanov:2019qjb,Bogdanov:2021yip}. Similarly, future gravitational wave observations will also enrich the neutron star mass-radius measurements~\cite{LIGOScientific:2018cki}. Mass or radius measurements performed separately can also be useful, since they can be used to rule out certain ranges of parameters or EOS, for example by comparing the maximum allowed mass in a theory to the most massive neutron stars that have been observed~\cite{Sotani:2017pfj}. Incorporating such data into our analysis can increase the precision of our results, and potentially enable us to effectively constrain spontaneous scalarization in the $\mphi \neq 0$ case as well.

We should note that our results are specific to the conformal scaling function $A(\phi) = e^{\beta \phi^2/2}$, which is most commonly used in the literature. The tachyonic instability that causes spontaneous scalarization is excited for any $A$ with a similar leading order behavior around  $\phi=0$, i.e. $A(\phi)= 1+\beta \phi^2/2 + \dots$. However, the ultimate structure of the neutron star, hence the mass-radius relationship, is determined by the nonlinear region of $A$. For example, $A_\infty \equiv A(\infty)$ controls the behavior of the theory at very large scalar amplitudes, hence, $1-A_\infty$ is a rough indicator of how much a highly scalarized object can deviate from GR. Consider the choice $A(\phi) = (1-\Delta) +  \Delta\ e^{\beta \phi^2/(2\Delta)}$ for $0 < \Delta < 1$. $\Delta \ll 1$ would ensure that deviations from GR are small even for high values of $\beta$, $\Delta=0$ simply being GR with a minimally coupled scalar. Ruling out such a theory using mass-radius data can be considerably harder since the mass-radius curves can be practically indistinguishable from those of GR within observational uncertainties. This discussion is not hypothetical, since the ``asymmetron'' mechanism of cosmology uses a similar conformal scaling~\cite{Chen:2015zmx}. Lastly, it is also possible to have a linear term as in $A(\phi)= 1+\alpha \phi + \beta \phi^2/2 + \dots$, which would not necessarily have GR configurations as a solution~\cite{Rosca-Mead:2020bzt}.

Generalizing the mass term to a generic self interaction, $\mphi^2 \phi^2/2 \to V(\phi)$, affects scalarization~\cite{Staykov:2018hhc}, which is another factor that can be considered in future studies. We should note that such terms can also have a role in the superradiant instability of scalar fields, which may result in bounds on their masses that are different from the ones we used in Sec.~\ref{sec:mphi_constraint}~\cite{Stott:2020gjj}. 

A shortcoming of our methodology is the simplifying approximations in comparing the current mass-radius data to theoretical structure of scalarized stars, as explained in Sec.~\ref{sec:mass_radius_curves}. We checked that this assumption provides the right order-of-magnitude results for the posterior distribution by repeating our calculation with two different approximations for the measured neutron star mass, which in some sense correspond to the two extreme options (see Eq.~\eqref{eq:mass_range}). Our main results about massless and massive scalars were robust within this range of approximations, as well as different choices of prior. Nevertheless, the quantitative differences between the approximations are significant, and a precise bound on the theory parameters would require a detailed reanalysis of the actual X-ray data for the scalarized neutron star spacetime. This, and adding the effects of neutron star spin will be future steps in refining our approach. 

Finally, we should emphasize that even though we studied the specific model of action~\eqref{st_action}, mass-radius data can be used to constrain a wide array of gravity theories. The only part of the procedure that changes is how neutron star structures are computed. Primary candidates for this approach are more recent spontaneous scalarization models that arise from general couplings between scalar fields and curvature terms~\cite{Doneva:2017bvd,Silva:2017uqg,Herdeiro:2018wub}. Whether the shortcomings of our analysis, especially regarding massive scalars, are still present in these theories to the same extend remains to be seen. We hope our results inspire further work in this direction.

\acknowledgments
We thank Roxana Rosca-Mead, Davide Gerosa and Ulrich Sperhake for their comments about solving the TOV equations as a boundary value problem, and David Marsh and Viraf Mehta for providing the details of the constraints on ultralight scalar masses from black hole superradiance. We also thank Andrew Coates for many valuable suggestions. K\.I\"U and FMR were supported by Grant No. 117F295 of the Scientific and Technological Research Council of Turkey (T\"{U}B\.{I}TAK). FMR was further supported by a Bilim Akademisi Young Scientist Award (BAGEP). We also acknowledge networking support by the GWverse COST Action CA16104, ``Black holes, gravitational waves and fundamental physics.''

\appendix
\section{Computation of neutron star structures}
\label{app:mass_radius}
A common way to solve the TOV-like equations in~\eqref{tov} is the shooting method~\cite{Ramazanoglu:2016kul}. One starts numerically integrating from the origin with a given central pressure $\pt_c$ and a guess value for central scalar field $\phi_c$, and updates $\phi_c$ according to the behavior of $\phi$ far from the star until the scalar is purely decaying within numerical accuracy. Even though this method works adequately for relatively small values of $(\beta,\mphi)$, it could not provide accurate solutions for the extremely massive neutron stars encountered for higher values of $|\beta|$. Instead, we cast Eq.~\eqref{tov} as a boundary value problem, discretize it using finite difference formulas, and solve the resulting nonlinear system using a relaxation method. This approach closely follows Ref.~\cite{Rosca-Mead:2020bzt} which can be consulted for details. We discuss below some of our efforts to reduce computational cost of this general approach.

Firstly, we observe that $\nu$ does not appear anywhere on the right hand side of Eq.~\eqref{tov}, hence we do not have to consider the $\nu'$ equation at all aside from inserting its right hand side to the equation for $\pt'$. We are ultimately interested in calculating $\nu$ to find the properties of the spacetime for a scalarized star, however this can be done by simply integrating the $\nu'$ equation once, after the relaxation process is over and we find the $\phi_c$ value that provides the correct physical behavior for $\phi(r \to \infty)$. If we include the $\nu'$ equations as one to be discretized and relaxed like the others in Eq.~\eqref{tov}, we increase the size of the Jacobian matrix blocks used in the relaxation method from $4\times8$ to $5\times10$ inside the neutron star, and $3\times6$ to $4\times8$ outside. Hence, ignoring the $\nu'$ equation during relaxation reduces the numerical cost by roughly a half.

Another potential problem of numerical error is the $\pt'$ equation near the surface of the star. The piecewise polytropes we investigated all behave as $\rt = C \pt^n$ around $\pt \approx 0$, where $0<n<1$. When the Jacobian is constructed for relaxation, $d\rt/d\pt$ brings divergent terms at the surface of the star where $\pt=0$. To our knowledge, this did not pose a problem in Ref.~\cite{Rosca-Mead:2020bzt}. Nevertheless, we
defined a new variable $\qt$ as
\begin{align}
d\qt = \frac{d\pt}{\pt+\rt(\pt)}\ \  , \  \qt(\pt=0)=0 \ .
\end{align}
This variable avoids divergences at the surface of the star, and also simplifies the equations to be solved. It can be easily calculated for any EOS $\rt(\pt)$, and can be expressed in terms of analytical formulas for polytropes or piecewise polytropes.

To sum up, the system of equations we solve is
\begin{align}\label{tov2}
 \mu' &= 4\pi r^2 A^4 \rt + \frac{1}{2}r(r-2\mu) \psi^2 + \frac{1}{2} r^2 \mphi^2 \phi^2 \nonumber \\
 \phi'&= \psi \nonumber \\
 \psi' &(r-2\mu) = 4\pi r A^4\left[\alpha(\rt-3\pt)+r\psi(\rt-\pt)\right] \\
 &\ \ \ \ \ \ \ \ \ \ \ \ \  +\mphi^2 (r^2\phi^2\psi +r\phi)-2\psi(1-\mu/r) \nonumber \\
 \qt' &= -\frac{1}{2}r\psi^2 
 - \frac{r^2}{2(r-2\mu)}\left[(8\pi A^4\pt-\mphi^2\phi^2)+2\frac{\mu}{r^3} \right] 
- \alpha \psi\ , \nonumber
\end{align}
where $\pt$ and $\rt$ are considered to be functions of $\qt$. The boundary conditions at the origin and infinity are
\begin{align}
\mu(r=0)=0\  , \psi(r=0) = 0 \ , \phi(r=\infty)=0\ .  
\end{align}
The variables inside and outside the star are also related to each other using internal boundary conditions at the surface of the star as in Ref.~\cite{Rosca-Mead:2020bzt}, however we set $\qt=0$ on the surface of the star, and do not have any matter-related variables outside.

The Jacobian of the linearized discretized equations~\eqref{tov2} has a block diagonal form that makes its solution much faster than the overall size of the matrix might suggest. Basically, individual blocks can  be solved in succession with a custom written code as explained in Ref.~\cite{Rosca-Mead:2020bzt}, which is adapted from Ref.~\cite{10.5555/1403886}. We followed an alternative method, and constructed a single large block diagonal Jacobian in the form of a sparse matrix in the Eigen library~\cite{eigenweb} at every iteration of the relaxation. We solved the linear system by using the \texttt{SparseLU} solver of Eigen. We checked the convergence of our solutions, and confirmed that they had the second order behavior consistent with our finite difference stencils.

\section{Computation of  mass-radius curves}
\label{sec:app_mr_curve}
The above process describes how to obtain a single scalarized neutron star solution, whereas we need mass-radius curves which require many such solutions for each EOS and point in the $(\beta, \mphi)$ parameter space. Thus, computation of all the curves and performing the likelihood calculation required computing many millions of individual stars. Doing this in a random manner can be computationally costly, thus, we followed certain guidelines to speed up the solutions.

A star is specified by its radius in the relaxation method we use, the same way as in Ref.~\cite{Rosca-Mead:2020bzt}. This is unlike the case of the shooting method where we specify a star in terms of it central pressure, or equivalently, density ($\rt_c$). Using the star radius is inconvenient since for a given EOS, all possible neutron stars form a one parameter family for the central density whereas this is not the case for the radius, i.e. in general there can be more than one neutron star with the same radius, which complicates the root finding procedure (see Fig.~\ref{mr_curve_beta_low}). Furthermore, we do not know in advance how large the scalar field would be for a given radius value, or if the star scalarizes at all. 

We followed some heuristic steps to calculate the mass-radius curves for a given point on the parameter space with as little computational waste as possible. To begin with, we remind that we can obtain the mass-radius curve in GR for any given EOS relatively easily, for example using the simpler shooting method. This curve forms a reference for each EOS, and is also the solution for the GR-equivalent region $\beta>\beta_c$ or $\mphi>m_c(\beta)$. 

We start with obtaining a single mass-radius curve for a given point on the parameter space, i.e. for a given EOS, $\beta$ and $\mphi$. It is known from previous studies that very low mass neutron stars do not scalarize~\cite{PhysRevLett.70.2220}. We start our construction of the mass-radius curve with a star that has very large radius, very low mass and very low $\rt_c$, which is known to be not scalarized, hence its structure is already known from the GR calculation. Previous studies also showed that stable scalarized stars are continuously connected to those of GR, hence the fist scalarized stars we encounter while increasing the value of $\rt_c$ are supposed to have low scalar field values~\cite{PhysRevLett.70.2220,Ramazanoglu:2016kul}.

\begin{figure*}
\centering
\includegraphics[width=\textwidth]{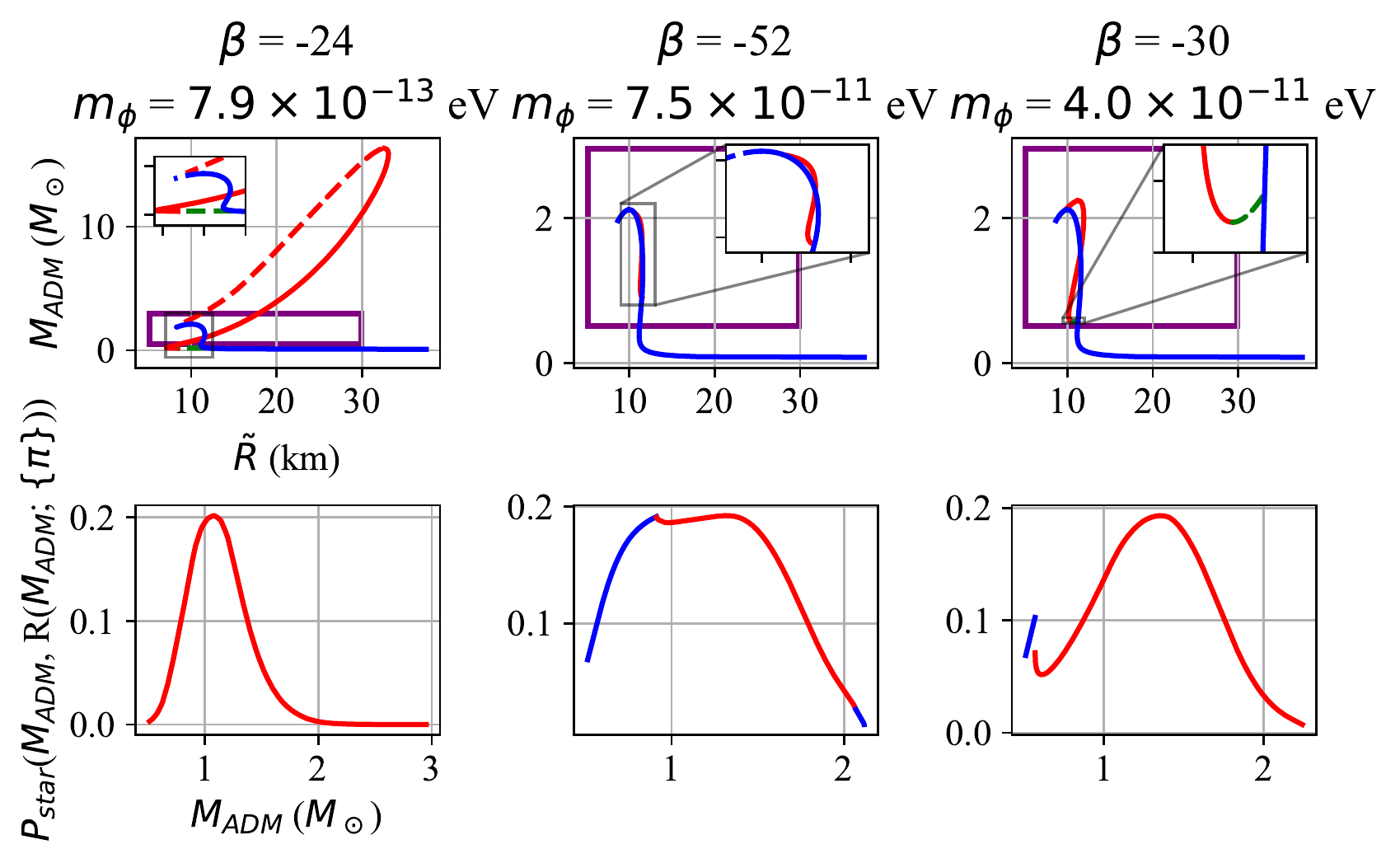}
\caption{Left column: $\beta=-24$, $\mphi=7.9\times10^{-13}$eV. Middle column: $\beta=-52$, $\mphi=7.5\times10^{-11}$eV. Right column: $\beta=-30$, $\mphi=4.0\times10^{-11}$eV. First row: Mass-radius curves for three different $(\beta, \mphi)$ pairs and HB EOS. Second row: The corresponding conditional probability densities $P_{4\textsf{U-} 1820 \textsf{-} 30} \left(M_{\rm ADM}, R(M_{\rm ADM}; \beta,\mphi,\lambda) \right) $ as defined inside the integral in Eq. \eqref{eq: distr of a star, given mass and params} for the neutron star $4\textsf{U-} 1820 \textsf{-} 30$ of Ref.~\cite{Ozel:2015fia}. Red curves show the scalarized stars, blue curves show the GR stars; solid lines are stable, dashed lines are unstable stars (they are excluded in the likelihood computation). Green dashed lines show the low-mass, unstable, scalarized stars, which were also excluded from the likelihood (also see Fig.~\ref{mr_curve_beta_low}). The purple box shows the domain in the mass-radius plane where the probability density of mass and radius of $4\textsf{U-} 1820 \textsf{-} 30$ is nonzero, only the pieces of mass-radius curves of the scalarized theory that lie inside the box contribute to the posterior likelihood. The box encloses the same region in the mass-radius plane in all three figures. For each of the $(\beta, \mphi)$ pairs, the contribution of the $4\textsf{U-} 1820 \textsf{-} 30$ data to the posterior is the integral of the corresponding curve in the second row. Left: An example of a highly scalarized theory where the intersection of the mass-radius curve with the data is completely scalarized. Middle: A weakly scalarized theory for which the likelihood includes GR contributions in both high and low masses. Right: The effect of excluding the low-mass unstable neutron stars for the $\beta = -30$ $\mphi = 4.0 \times 10^{-11}$~eV theory: It leads to a discontinuity in the corresponding conditional probability. This discontinuity has a negligible effect since it concerns a short interval of very low neutron star masses where the corresponding conditional probability is low.
}
\label{fig:mr_curve_parts}
\end{figure*}
After we start at a very low $\rt_c$ value where there is no scalarization, we aim to monotonically increase $\rt_c$ with a given approximate step size $\Delta \rt_c$. The challenge is that we can only directly specify the radius $R$ in our relaxation code, not $\rt_c$. To overcome this, we use the $R$ and $\rt_c$ information of the last few stars we already computed to estimate the radius of the next star on the curve we aim to compute. For example, assume that the last stars we computed had approximate central densities $\rt_l$, $\rt_l- \Delta \rt_c$, $\rt_l- 2\Delta \rt_c$, and corresponding radii. We use Lagrange polynomials on the function $R(\rt_c)$ based on the already computed stars to extrapolate the radius of the next star, $R(\rt_l +\Delta \rt)$, and perform relaxation for this radius value. This relaxation typically provides a solution where the central density is close to $\rt_l +\Delta \rt$. We also use the same extrapolation strategy to obtain very accurate initial guesses for the density and scalar field profiles for this star, which drastically decreases the number of relaxation cycles, thusly also the computation time. We then repeat the process for the next higher $\rt_c$ value.

Note that extrapolation can also be performed in pieces of the mass-radius curve if we use $R$ to parametrize it, however we cannot cover all the mass-radius curve from one end to the other by incrementally changing $R$. We jump to different parts of the curve or between the scalarized and GR branches when $M(R)$ becomes multivalued. The jumping behavior on the mass-radius curve means that we sometimes cannot reach certain physically relevant parts of it. This led us to the more reliable $\rt_c$ parametrization scheme which also requires considerably less manual correction, an important criterion since we needed $~2\times 10^4$ mass-radius curves for our analysis, one for each point in the parameter space.

We also use an adaptive $\Delta \rt$ to make sure we compute the curve with sufficient resolution. This is especially important for high values of $\beta$ where the mass and radius of a star can change rapidly with $\rt_c$. 

The mass-radius curve obtained from the previous steps is finally further restricted to the stable stars, which are the only ones that are astrophysically relevant. The methodology can be seen in Fig.~\ref{fig:mr_curve_parts} for different scenarios. Our main criterion for stability is  
\begin{align}
\frac{dM_\text{ADM}}{d\rt_c}>0 \ .
\end{align}
Stars on the parts of the curve where this criterion is not satisfied, such as those to the left of the maximum mass in Fig.~\ref{mr_curve_beta_low}, are known to be unstable in GR~\cite{Shapiro:1983du}. This was also observed to be the case for spontaneous scalarization so far. Once the highest ADM mass is reached in the mass-radius curve, we still obtain scalarized solutions with even higher values of $\rt_c$, but lower ADM masses, however, these are unstable as in GR~\cite{Ramazanoglu:2016kul}.

We discovered a second region of instability in the mass-radius curves of scalarized stars for large values of $|\beta|$. Namely, the scalarized stars with the lowest $\rt_c$ values where the mass-radius curve for spontaneous scalarization branches off from that of GR for the first time are also found to be unstable by the above criterion (see Fig.~\ref{mr_curve_beta_low}). For any star under this category, we confirmed that there are other scalarized stars with the same baryon mass, but with a higher binding energy, the difference between ADM and baryon masses, which is known to be a strong indication for instability~\cite{Mendes:2016fby}. This new category of unstable stars also mean that stable stars may not be able to have arbitrarily weak scalar field clouds if $\beta \lesssim -10$, depending on $\mphi$. This might have interesting astrophysical implications on its own. Regardless of all the instability arguments so far, the last category of stars are low-mass, and their overlap with the mass-radius data is tiny. Therefore, the contribution of this region of the mass-radius curves to our posterior likelihood is insignificant.

To summarize, we use the stable portions of the mass-radius curve in the likelihood calculations, which means the GR mass-radius curve is used for the neutron star mass values where there is no stable scalarized star. Due to the instability of the most weakly scalarized stars we mentioned above, these curves are sometimes discontinuous for high $|\beta|$, but the region of discontinuity has a very small contribution to the likelihood since it occurs at a very low neutron star mass (see Fig.~\ref{fig:mr_curve_parts}).

Finally, we repeat this process for each point of the parameter space. Scalarization characteristics are continuous with respect to $\beta$ and $\mphi$, hence we expect neighboring points on the $(\beta,\mphi)$ plane to have very similar mass-radius curves. Thus, we use the information for the already computed curves when we move to a new neighboring $(\beta,\mphi)$ point in order to speed up the curve construction. 

\section{Interpolation of data}
\label{sec:interpolation}
\begin{figure}
\centering
\includegraphics[width=0.49\textwidth]{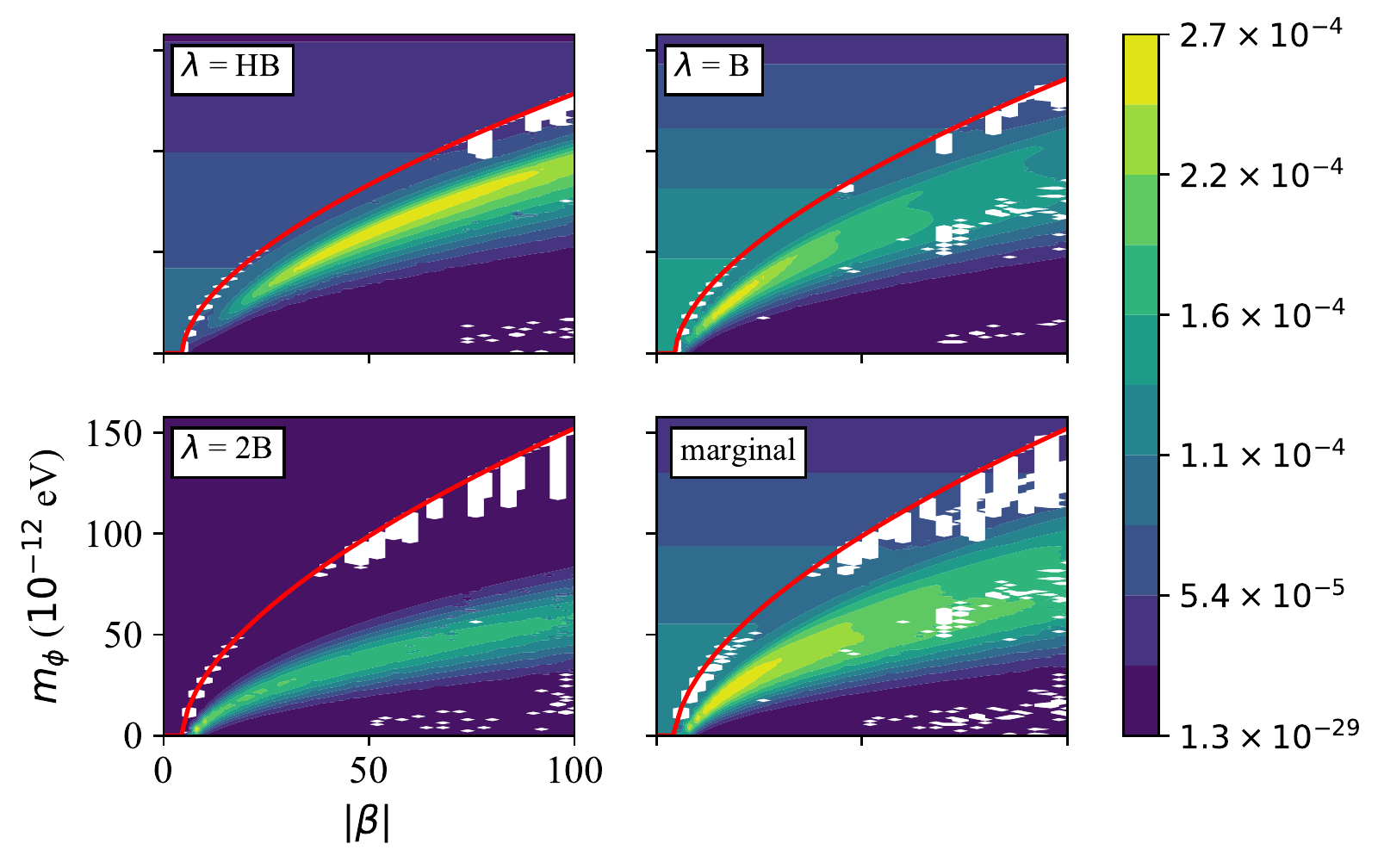}
    \caption{Posterior likelihoods for the mass-cutoff prior before smoothing and interpolation. The white regions and spots are where our code fails to compute the mass-radius curves, i.e., some or all of the neutron star mass-radius curve that intersects the observational data is missing. Scattered spots concentrated at high $|\beta|$ and usually low $\mphi$ are extremely scalarized cases. Larger empty regions near the $m_c(\beta)$ curve correspond to weak scalarization where the code failed to find the scalarized branch of the mass-radius curve. These weakly scalarized stars are almost identical to those of GR.
    } 
\label{fig:unprocessed_posteriors_2D}
\end{figure}
\begin{figure}
\centering
\includegraphics[width=0.49\textwidth]{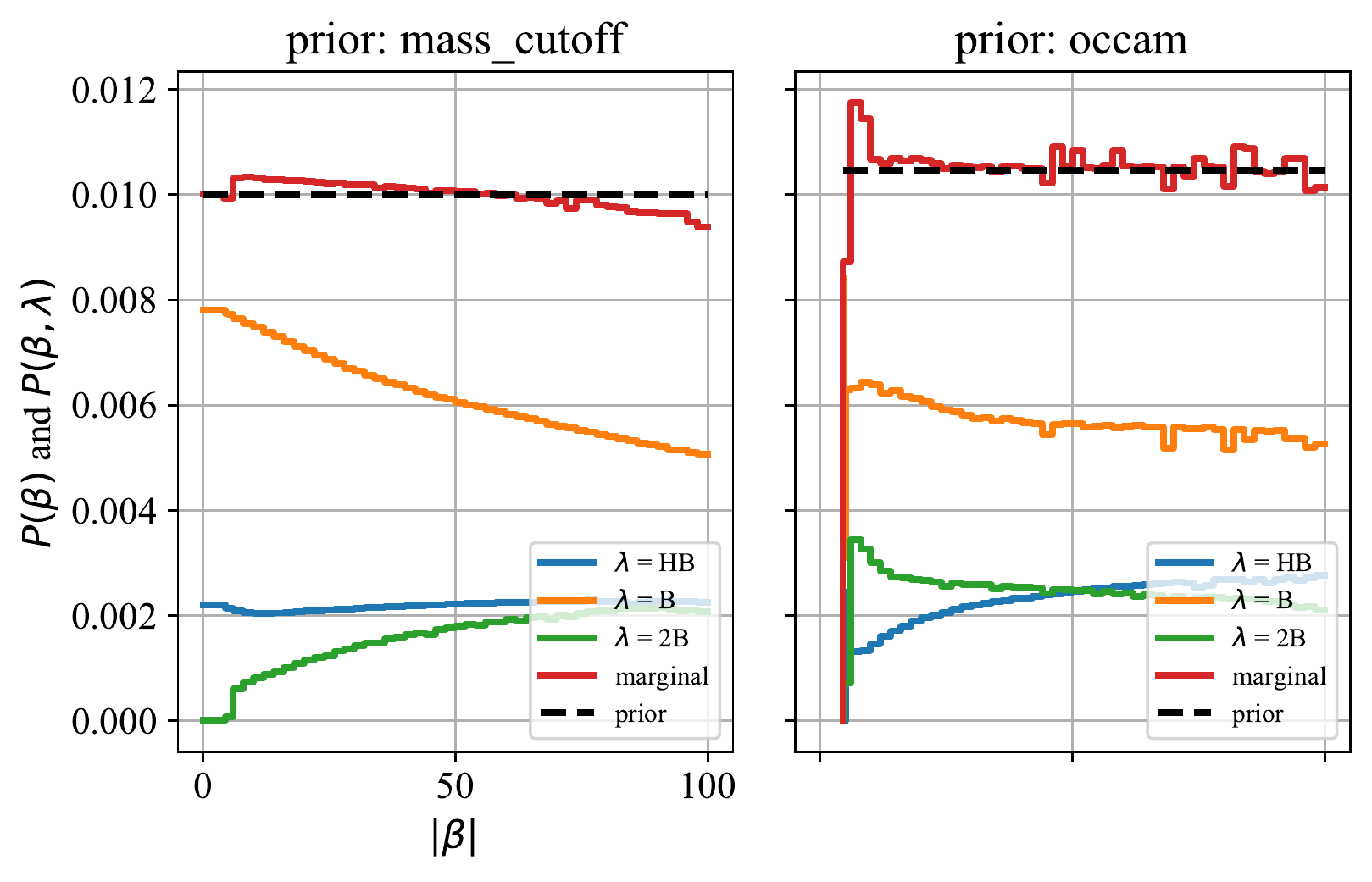}
    \caption{The conditional and marginal posteriors obtained from Fig.~\ref{fig:unprocessed_posteriors_2D} without interpolation. Occam's prior is most affected by the missing points near the $m_c(\beta)$ curve, especially at high $\beta$ values. However, the behavior is the same as the interpolated data in Fig.~\ref{beta_marginal_massive}, and the bounds on $\beta$, or rather, our inability to obtain one, are not affected.
    } 
\label{fig:unprocessed_posteriors_1D}
\end{figure}
\begin{figure}
\centering
\includegraphics[width=0.49\textwidth]{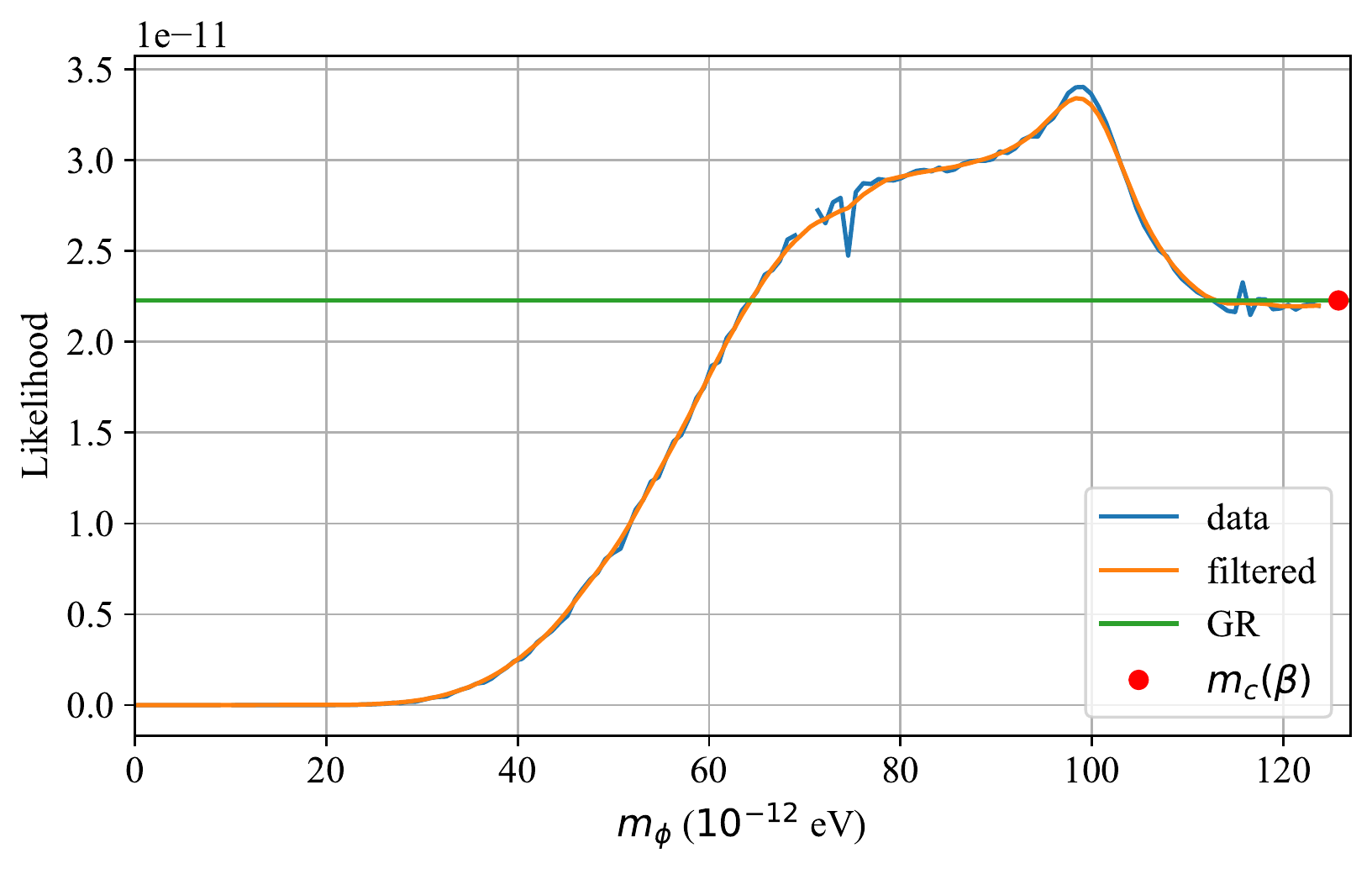}
    \caption{An example of the filtering and interpolating process described in the text. The raw values of the likelihood function along the $\beta = -88$ slice for B EOS (data, blue), the output of the filtering and interpolation (filtered, orange) and the GR likelihood value (GR, green) are plotted together. There are regions where filtering is not required at all, and some where the raw data has more noise and spikes. Filtering slightly changes the maximum of the likelihood, however, the error introduced by this is much smaller than the uncertainty arising from other sources such as the prior choice.} 
\label{fig:unprocessed_posteriors_fixed_beta}
\end{figure}

Since we have a discrete grid, we cannot know the exact location of the $m_c(\beta)$ curve, which is an important factor, especially in determining Occam's prior. Moreover, our relaxation scheme failed to find scalarized solutions for certain parameter values, which meant we could not construct part or whole of the mass-radius curve. We used interpolation from nearby regions on the parameter space to address both of these issues.

The failures of our relaxation method can be seen in Fig.~\ref{fig:unprocessed_posteriors_2D} in white. The ``unprocessed'' posterior likelihoods on $\beta$ alone (obtained by marginalizing over $\mphi$ in Fig.~\ref{fig:unprocessed_posteriors_2D}) can be seen in Fig.~\ref{fig:unprocessed_posteriors_1D}. There are two main regions of failure. First, there are scattered spots in high $\beta$ and typically, but not always, low $\mphi$ values where scalarization is high (lower right white spots in Fig.~\ref{fig:unprocessed_posteriors_2D}). Second, there are wider continuous regions near the $m_c(\beta)$ curve where scalarization is quite low.

Scalarization depends continuously on $(\beta,\mphi)$, hence filling the empty individual spots of the first region using interpolation from their neighbors is relatively simple and safe. The second region is continuous, meaning nearest or next-nearest neighbors of points also failed in relaxation, and interpolation is not straightforward. Even though it might seem at first that it would be harder to determine the fate of such points, understanding why they occur makes addressing this problem relatively easy. Our code struggles here to find the scalarized branch of the mass-radius curve, because the curve is almost indistinguishable from that of GR due to very weak scalarization. This is confirmed by checking the curves at nearby points of the parameter space where relaxation was successful. We still try interpolation in this region, but, as explained in more detail below, we sometimes simply use the GR mass-radius curves without introducing any appreciable error.

One of the first tasks in the posterior calculation is locating the GR-equivalent region to determine where to actually look for scalarization, and where to simply use GR results. After sampling $\sim 10^3$ points on the $(\beta, \mphi)$ parameter space for a given EOS, we first construct the $m_c(\beta)$ curve as follows: For each value of $\beta \in [-100, 0]$, we find the value of $\mphi$ for which the likelihood function is closest to the GR likelihood value. This corresponds to the highest value of $\mphi$ at which our code could find a scalarized branch on the mass-radius curve for this $\beta$ value. We know that $m_c(\beta)$ is a monotone function of $-\beta$, although for some points in the parameter space, our code failed finding scalarized mass-radius curves beyond the $\mphi$ value of a larger $\beta$ value for which it found scalarization. Hence, we sample a monotone subset of the $m_c$ values we obtain from our likelihood and make the functional fit
\begin{align}
    m_c(\beta) = a(-\beta - b)^c
\end{align}
with parameters $a,b,c$ to this data. After obtaining the $m_c(\beta)$ curve, we declare the portion of the parameter space that lies above $m_c(\beta)$ to be the GR-equivalent region where the mass-radius curves are those of GR.

The interpolation of the likelihood function is challenging not only due to the regions on the $(\beta,\mphi)$ plane where relaxation fails, but also due to the fact that posterior has a ``noisy'' pattern in some cases, see Fig.~\ref{fig:unprocessed_posteriors_fixed_beta}. The noise is mainly due to the discrete nature of the probability distributions of the stars on their mass-radius space, $P_i(M,R)$, which causes the intersection of the theoretical curves with these distributions to change in sudden jumps as the curves change with $(\beta,\mphi)$ (see Sec.~\ref{sec:bayes}). Such noise can be further enhanced when further numerical interpolations and numerical integrations are performed, such as when marginalizing over $\mphi$ in the posterior. Because of this, we used smoothing on the final likelihood data. Specifically, we used the \texttt{RectBivariateSpline} method with degree 1 of the SciPy module of Python. Since the likelihood function was not smooth enough, we avoided 2D interpolation. Instead, we took the constant $\beta$ lines of the likelihood function as one-dimensional curves and smoothed them at the regions where they were noisy using a smoothing filter.

When necessary, we interpolated the missing points on the grid arising from the failure of relaxation using the data filtered and smoothed as above. For the high $\mphi$ values just below the $m_c(\beta)$ curve, the interpolation does not always give accurate results, mainly due to the residual noise in the data, even after the filtering. We substituted the GR likelihood value for the value of the likelihood function at these regions, which was tolerable since the mass-radius curves in this region are already very weakly scalarized, making the likelihood function close to the GR value.

We used the simple trapezoid method for the integrals performed in the computation of marginal and conditional probabilities. Even though this method has a relatively low degree of convergence, the truncation error it causes is well below the uncertainty from other sources, and it behaves robustly on residual noise on the integrand, compared to higher-order converging integration schemes.

\bibliography{references_all}

\end{document}